\documentclass[journal]{IEEEtran}
\usepackage[shortcuts]{extdash}
\usepackage{bm,upgreek}

\usepackage{graphicx}
\ifCLASSINFOpdf
\else
\fi
\usepackage{amsmath}
\usepackage{amssymb}
\usepackage{amsfonts}

\usepackage{upgreek}

\begin{document}
%
\title{Quantized Signal Recovery with Interference via Parametrized Look-Up Tables}
%
%
%

\author{Morriel~Kasher,~\IEEEmembership{Student~Member,~IEEE,}
        Michael~Tinston,~\IEEEmembership{Member,~IEEE,}
        and~Predrag~Spasojevic,~\IEEEmembership{Senior~Member,~IEEE}

\thanks{This material is based upon work supported by the Office of Naval Research under Contract No. N68335-21-C-0625 and the National Science Foundation Graduate Research Fellowship under Grant No. DGE-2233066.}%
\thanks{M. Kasher and P. Spasojevic are with the Department of Electrical and Computer Engineering, Rutgers University, New Brunswick, NJ 08901 USA (email: morriel.kasher@rutgers.edu; spasojev@winlab.rutgers.edu)}%
\thanks{M. Tinston is with Expedition Technology, Inc., Herndon, VA 20171 USA (email: mike.tinston@exptechinc.com)}%
}

%
%

\markboth{Preprint}%
{Kasher \MakeLowercase{\textit{et al.}}: Quantized Signal Recovery with Interference via Parametrized Look-Up Tables}
%



\maketitle

\begin{abstract}
Efficient all-digital post-correction of low-resolution analog-to-digital converters can be achieved by using Look-Up Tables (LUTs). The performance of a LUT can be optimized by incorporating a parametric model for the expected input signal, noise level, and interference signals. We evaluate three analytical estimators for integration with parametrized LUTs, especially with applications to low-resolution, non-linear, or wideband quantizers. We also propose several approximations to improve tractability of the estimation problem for Phase-Shift Keyed input signals and Linear Frequency Modulated interference signals. Simulated results validate the ability of our estimator to recover the instantaneous value of the desired input signal in real-time with a high degree of accuracy. This includes cancellation of harmonic distortion that aliases into the desired signal bandwidth from front-end saturation due to high-power out-of-band interference. Our estimators are shown to achieve a significant gain over conventional linear-filtering techniques while also being robust to changes in input parameters, non-linear quantizers, and time-variant interference sources. For a tone input quantized to 3 bits and estimated with a fixed 12-tap model order we achieve $>$10 dB improvement in Mean Square Error and $>$20 dBc improvement in Spurious-Free Dynamic Range.
\end{abstract}


%
\IEEEpeerreviewmaketitle

\section{Introduction}
%
%
%
%
\IEEEPARstart{Q}{uantization} 
(or analog-to-digital conversion) is a ubiquitous process in audio/video, measurement, data compression, and communication systems. A quantizer/analog-to-digital converter (ADC) applies a hard non-linearity to a continuous-domain analog input to produce a discrete-valued output. While this operation is necessary to represent signals digitally or reduce their size, it introduces quantization error which distorts the input signal and limits the accuracy of its digital representation \cite{GrayQuant1998}.

Conventional quantization produces an error process that is highly (self-)correlated and also correlated with the input signal, resulting in prominent quantization artifacts that can reduce the perceived fidelity of quantized data~\cite{GrayDither1993}. This effect is especially significant in low-resolution quantization required for low-latency wideband applications and, hence, motivates a method to decorrelate quantization error.
The non-subtractive dithering method achieves this by adding an analog dither signal to the input prior to quantization \cite{wannamaker1997thesis}. When drawn from an appropriately-chosen distribution, the dither signal can render conditional moments of the error process independent of the input \cite{WannamakerNSD2000} \cite{vanderkooy1984resolution}. This has been shown to significantly improve perceptual quality of the quantized output \cite{WannamakerPsychoacoustics1992}.

While dithering is an attractive and effective option for decorrelating quantization error, it can be challenging to implement in practice. This is because it requires real-time generation and addition of the 
 analog dither  signal 
prior to analog-to-digital conversion, typically necessitating an entire parallel digital-to-analog converter (DAC) chain. 
Here, we consider a method for compensating ADC quantization error exclusively in the digital-domain (i.e., post-quantization). 

A Look-Up Table (LUT) is an all-digital post-processing method used to improve quantizer performance efficiently. A state-space-indexing LUT uses $N$ previous quantized output values to index a correction value which then replaces the current digital output \cite{lundinthesis} \cite{lundin_characterization_2005}. We call $N$ the order or dimensionality of the LUT. This structure can be implemented using  two-level combinational logic circuits having $\mathcal{O}(1)$ processing time complexity, 
making it well-suited for wideband applications due to its extremely low latency. Furthermore, as an all-digital method it can be 
integrated with an existing digital signal processing back-end.

Despite promising characteristics, existing work on LUT design is limited. The choice of LUT output value is non-trivial. Prior work addresses how to optimize it for Mean Square Error (MSE) \cite{adctextbook}, Total Harmonic Distortion (THD) \cite{lundin2003minimalthlut} \cite{hummels1} \cite{hummels2}, and Spurious-Free Dynamic Range (SFDR) \cite{kashersfdrlut}. 
However, these works lack a set of cohesive LUT design principles--- existing literature relies almost exclusively on data-driven calibration procedures to train and optimize the LUT entries numerically \cite{hummels1} \cite{de_vito_bayesian_2007} \cite{lundin_external_2001} \cite{gines_digital_2021}.
These design methods are unreliable (subject to the quality of training data), uninformative (functionally black-box approaches), and have unpredictable performance (unable to be studied analytically).
The only analytically-derived LUTs are 
the popular Midpoint and MMSE LUTs, designed only for $N=1$ \cite{adctextbook} \cite{attivissimomidpointlinearizationNSdither} \cite{lundin_bounds_2009}. These works restrict LUTs to compensating ADC irregularities (e.g., for a non-uniform transfer function), and they typically require high-precision outputs, whereas, they suffer severe performance penalties when their resolution is limited to that of the input \cite{lundin_adc_2005}.
Another important research direction are parametrized LUTs. One popular example are the frequency-selective LUTs \cite{lundin_analog--digital_2002}, where the input signal is assumed to be a  tone and its  frequency is estimated using  an additional LUT \cite{andersson_frequency_2000} \cite{kasherfreqestlut}.
Moreover, all existing LUT-based corrections require large memory size that grows exponentially with $N$. This makes LUTs impractical for memory-constrained systems such as Field Programmable Gate Arrays (FPGAs) and Integrated Circuits (ICs).

To overcome these limitations, we study a broad class of all-digital model-based  LUTs which we term \textit{dithered parametrized look-up tables}. 
Our design is decomposed into indexing, estimation, dithering, and re-quantization stages. 
Prior information about the input signal informs its parameters estimated by a LUT indexed using quantized low resolution ADC samples. 
Several strategies emulate the desirable effects of dithering in the digital domain, an approach we term {\em post-quantization dithering}. 
They allow the designer to trade MSE with SFDR via digital randomization, which reduces error correlation and improves spectral purity  at the cost of increased noise power~\cite{kasher_postquantization_2024}. Re-quantization to low/original resolution ensures that the low-latency and wideband post-processing are still feasible 
while maintaining performance. 
However, successful implementation of this novel LUT architecture relies on the derivation of an accurate estimator for each incoming sample in real-time relying only on previous quantized observations. 

Such an estimator is constrained in several ways: it must be causal, deterministic, and insensitive to mismatch between theoretical and actual input signal. 
To use our LUTs in wideband receivers (spectrum analyzers or communication devices), our estimator must also account for simultaneous reception of high-power interfering signals alongside thermal noise. 
Moreover, the exponential memory requirement of our LUT architecture motivates its use with low-resolution quantizers. This further complicates the estimation task by preventing typical high-resolution quantization noise approximations like pseudo-quantization noise \cite{widrow_statistical_1996} or the additive noise model \cite{marco_validity_2005}.
Computational efficiency of the estimator is largely irrelevant since each estimate will be pre-computed for storage in a static LUT (with $\mathcal{O}(1)$ access time complexity).

To address these unique design constraints we improve upon several prior works in the literature on quantized estimation. The field of compressed sensing studies quantized parameter estimation in detail \cite{dirksen_quantized_2019} \cite{boche_quantization_2015} \cite{zymnis_compressed_2010}. This lends itself to Cramer-Rao Lower Bound (CRLB) derivation, a topic of several papers ranging from one-bit single-tone \cite{host-madsen_effects_2000}, multi-bit single-tone \cite{moschitta_cramerrao_2007}, and multi-bit multi-tone estimation \cite{fu_quantized_2018}. There also exists a general form for the Cramer-Rao Bound on parameter estimation from quantized measurements \cite{stoica_cramerrao_2022}. Work in the field of low-resolution communication has largely focused on channel estimation from quantized measurements, either for parameters of a non-linear channel \cite{mezghani_transmit_2009}, or for the impulse response of a linear channel quantized to one-bit \cite{ding_optimality_2024} \cite{fesl_linear_2024}. Some bounds exist on performance of M-PSK communication signals when quantized at baseband \cite{gayan_low-resolution_2020}. This was expanded to the computation of the Bayesian Cramer-Rao Bound (BCRB) for massive MIMO-OFDM systems in low-resolution \cite{thoota_massive_2022}.

All of these works share several shortcomings. Most 
restrict the input signal to be Gaussian or stationary while others restrict the estimator to be linear. 
Those papers which do not rely on these conditions consistently assume conditional independence of the quantized observations when conditioned on the parameter being estimated. 
By contrast, our work seeks to estimate the instantaneous input sample which makes all quantized measurements conditionally dependent on the estimand. This is estimation problem is far more complicated and intractable, where the optimal estimator can achieve a significant gain over the naive parameter estimators proposed in the literature. We evaluate each estimator in the context of realistic spectral scenarios in communication environments or measurement systems rather than simply 
bounding their statistical efficiency (via CRLB). 
Our approach is particularly practical for oversampling receivers due to the 
inclusion of interference signals (a topic that is unstudied by the aforementioned literature) and our study of wideband input signals (such as Phase-Shift Keyed sinusoids).

The primary contribution of our work is the proposal and evaluation of several estimators for digitally recovering quantized signals to be used in a parametrized look-up table architecture. 
This contribution is unique in that it combines several distinct features and advantages:
\begin{itemize}
    \item Signal recovery after low-resolution quantization, a particularly difficult challenge due to the introduction of strong deterministic non-linear distortion that cannot be well-approximated by any conventional stochastic noise models.
    \item Generalization to recovery of arbitrary input signals; substituting reliance on highly-restrictive signal assumptions (such as Gaussian or stationary inputs); instead, using Bayesian prior distributions describing a parametric model for the expected signal, noise, and interference at the quantizer input. 
    \item Robustness to the presence of noise at the input signal prior to quantization and ability to effectively perform in-band  denoising.
    \item Robustness to the presence of interference signals at the input prior to quantization, including high-power interference, saturation, and in-band interference. The latter two issues are uniquely difficult to correct since they are immune to traditional linear-filtering techniques, necessitating our non-linear recovery approach.
    \item Versatility to arbitrarily non-linear quantizers, which further augment the signal model beyond the classic uniform quantization effects.
    \item All the benefits of LUT-based post-correction methods, including: all-digital implementation, $\mathcal{O}(1)$ access time complexity, two-level combinational logic for ultra-low latency, and pre-computation of entry values allowing training algorithms of arbitrary computational intensity without impact on run-time performance.
\end{itemize}
We also describe several approximations for tone-like inputs that allow a LUT 
to be applied to wideband phase-shift keyed or time-variant (linear frequency-modulated) inputs. Moreover, we propose a standardized set of test input signal combinations intended to exercise three particularly difficult low-resolution effects: in-band distortion of the signal of interest, in-band aliasing of harmonics from high-power out-of-band interference, and saturation (clipping) of the signal due to high total input power. Exemplary results include: improving a sinusoid's MSE by 10+ dB and SFDR by 20+ dBc using a 12th-order LUT, reducing MSE of any arbitrary frequency sinusoid by 3 dB using a single 8th-order LUT, improving EVM of a low-power BPSK waveform by 6+ dB, improving SFDR by 25+ dBc of a sinusoid received with high-power sinusoidal interference, and allowing successful frequency synchronization at 4 dB higher interference power using an 8th-order LUT.

\section{Preliminaries}
\subsection{Notation}
Table~\ref{tbl:notationtable} defines notation used throughout the paper.
Not included in the table are constants, which can have arbitrary capitalization and subscripts but are explicitly stated to be constants (ex: $K$, $N$, 
and sometimes $a,b$). Moreover, function definitions ($Q_{b}, \mathsf{\Pi}_{\Delta}$), and set definitions ($\mathcal{I}_{b}$) 
use subscripts to denote parameters instead of time.
\begin{table}[h]
    \vspace{-4mm}
    \caption{Notation} \label{tbl:notationtable}
    \vspace{-4mm}
    \begin{center}
        \begin{tabular}{ |c|c| }
            \hline
            \textbf{Style} & \textbf{Interpretation} \\
            \hline
            Uppercase ($X$) & Random Variable (R.V.) \\ 
            Lowercase ($x$) & Realization of R.V. \\ 
            Bold Uppercase ($\mathbf{X}$) & Vector-Valued R.V. or Matrix \\
            Bold Lowercase ($\mathbf{x}$) & Vector \\ 
            Hat ($\hat{x}$) & Estimate \\
            Tilde ($\,\widetilde{x}\,$) & Model/Approximation \\
            Calligraphic ($\mathcal{X}$) & Set or Transformation \\
            Subscript ($x_{n}$) & Time-Index ($x$ at time $n$) \\
            Bracketed Subscript ($x_{[i]}$) & Vector-Index ($i$-th element of $\mathbf{x}$) \\
            Text Subscript ($x_{\mathrm{TEXT}}$) & Property of Variable\\ 
            \hline
        \end{tabular}
    \end{center}
    \vspace{-4mm}
\end{table}


For convenience we define the rectangular function: 
\begin{equation}
    \mathsf{\Pi}_{a}(t) \triangleq 
    \left\{\begin{matrix}
    \frac{1}{a},& -\frac{a}{2} \leq t \leq \frac{a}{2}\\ 
    0,& \textup{otherwise}
    \end{matrix}\right.
\end{equation}
Note that for arbitrary $a,b$ we have $\mathsf{\Pi}_{(b-a)}(t-\frac{a+b}{2}) = \{ 1/(b-a), t \in [a, b]; 0, t \notin [a,b]$.

We denote the probability distribution of a random variable $X$ 
with $p_{X}(x) = p(X=x)$. When written without the subscript, the random variable is implied 
(ex: $p(y|x) = p_{Y|X}(y | x) = p(Y=y| X=x)$). For a continuous R.V., $W \sim f_{W}(w)$ where $f_{W}(w)$ is its probability density function.

\subsection{Quantization \& Dithering}
The scalar quantization operation $Q(.)$ is defined on inputs $x \in \mathbb{R}$ with unique monotonically-increasing digital codebook in a vector $\bf{C}$ and unique monotonically-increasing analog partition levels in a threshold vector $\bf{T}$ such that:
\begin{equation}\label{eq:quantizerdefn}
    Q(x) = C_{k},\;\; T_{k} < x < T_{k+1}
\end{equation}
where $k = \{1, \hdots, 2^{b}\}$ for a $b$-bit quantizer.
By convention the first partition value $T_{1} = -\infty$ and the last partition value $T_{2^{b}} = \infty$, ensuring $\mathrm{dom}(Q) = \mathbb{R}$.
Edge cases are handled by modeling $Q(T_k)$ as a stochastic variable realizing $C_{k}$ or $C_{k-1}$ each with 
probability $1/2$, emulating a meta-stable comparator state. 
A quantizer is \textit{uniform} if $C_{k+1}-C_{k} = T_{k+1}-T_{k} \triangleq \Delta$ for $k = \left\{2, \hdots, 2^{b}-1\right\}$. An infinite ($k \in \mathbb{Z}$) uniform quantizer can either be \textit{mid-tread} ($C_{k} = k\Delta, T_{k} = k\Delta - \Delta/2$) or \textit{mid-riser} ($C_{k} = k\Delta + \Delta/2, T_{k} = k\Delta$). In this paper we consider mid-riser quantizers since $T_{0} = 0$ ensures no dead-zone and, hence, the ability to represent arbitrarily low-amplitude signals. Furthermore, the $b$-bit uniform quantizers in this paper are normalized such that $\Delta = 2^{-b+1}$.

Denote the Integral Non-Linearity (INL) of a quantizer as $\mathrm{INL}_{k} \triangleq T_{k} - ((k-1)\Delta - 1), k \in \{2, \cdots, 2^{b}-1\}$. 
A quantizer corrupted by INL has a transfer function that deviates from the ideal uniform one, requiring unique treatment for correction.

A \textit{dithered} quantizer forms the input as a sum of an analog input sample $x$ and an additive dither sample $w$. When used non-subtractively, the output is $y = Q(x+w)$ which maintains the same fixed-point resolution as the undithered quantization. 
Dither can be an inherent property of the system, such as 
Gaussian noise $W \sim \mathcal{N}(0, \sigma^{2})$ prior to quantization due to thermal effects.
Alternatively, dither can be purposely added according to a particular distribution such as the popular uniform/rectangular dither $W \sim \mathrm{Uniform}(-\Delta/2, \Delta/2)$.
Rectangular dither ensures that the quantizer is asymptotically unbiased (mean absolute error converges to 0 when averaging samples) with uncorrelated quantization error 
\cite{wannamaker1997thesis} \cite{lipshitzdithersurvey}.

\subsection{Figures of Merit}
Designing a digital post-correction scheme requires a metric or optimization objective for evaluation and comparison of any proposed methods. In practice, desired performance may be difficult to quantify (e.g., perceptual prominence of quantization artifacts) or conflict with alternative objectives (e.g., ease of implementation in a practical system). 
Here, we describe four LUT evaluation metrics.

Mean Square Error (MSE) is a classic and straightforward metric defined for a desired reference signal $\mathbf{x}$ and a test signal $\mathbf{\hat{x}}$ as:
\begin{equation}
    \mathrm{MSE\; [dB]} \triangleq 10 \log_{10}\left( \mathbb{E}\left[\left(\mathbf{x} - \mathbf{\hat{x}}\right)^{2}\right] \right)
\end{equation}

Spurious Free Dynamic Range (SFDR) is a frequency-domain metric intended to better represent the perceptual impact of our quantizer. This is important as the use of dithering strictly increases MSE, but can imbue the output signal with many desirable statistical properties such as uncorrelated and spectrally white error which are not captured by MSE alone. It is defined only for sinusoidal input signals with known fixed frequency $\tilde{f}$ as:
\begin{equation}\label{eq:sfdrdefn}
    \mathrm{SFDR\; [dBc]} \triangleq 10 \log_{10} \left( \frac{ \left|\hat{X}(\tilde{f}\,)\right|^{2}} {{\displaystyle \max_{f \notin [\tilde{f} - f_{\mathrm{o}}, \tilde{f} + f_{\mathrm{o}}]}} \left|\hat{X}(f)\right|^{2}}\right)
\end{equation}
where $\hat{X}(f) = \mathcal{F}\{\mathbf{\hat{x}}\} = \sum_{n} \hat{x}_{n} \exp(-j 2\pi f n)$ is used to estimate the Power Spectral Density (PSD) of the signal by computing its periodogram as $|\hat{X}(f)|^{2}$ and $f_{\mathrm{o}}$ is an offset term.
This offset term is necessary because SFDR is computed for a sequence of samples whose finite length will generate sidelobes due to windowing and generate spectral leakage due to non-integer period. 
An example application where this criteria is critical is spectrum sensing or analysis, where the SFDR represents the maximum reliable dynamic range not containing quantization artifacts (``spurs'') which are detailed in Sec.~\ref{sec:motivation}.


In communication applications, Error Vector Magnitude (EVM) is a key metric due to its ability to quantify the likelihood of Bit Error Rate (BER) and hence the reliability of a communication link. It is defined for a recovered symbol $\hat{s}$ and a reference symbol $s$ as:
\begin{equation}
    \mathrm{EVM\; [dB]} = 10 \log_{10}\left( \frac{|\hat{s} - s|^{2}}{\mathbb{E}[|s|^{2}]} \right)
\end{equation}
where the the received symbols in a PSK-modulation scheme first undergo filtering, mixing, and sampling to recover the baseband constellation prior to EVM computation. For a set of symbols $\mathbf{s}$ the overall constellation has a root-mean-square EVM value defined similarly as $10 \log_{10}\left( \mathbb{E}[|\mathbf{\hat{s}} - \mathbf{s}|^{2}]/\mathbb{E}[|\mathbf{s}|^{2}] \right)$.


One use case for a LUT is to aid in coarse synchronization, namely Carrier Frequency Offset (CFO) correction. A common low-cost method to do this (relevant for latency-constrained wideband systems) described in \cite{sklar2001digital} is to estimate the instantaneous CFO of a received Binary Phase-Shift Keyed waveform as:
\begin{equation}
    \hat{f} = \frac{1}{2}\arg \max |\Psi(f)|\label{eq:CFOest}
\end{equation}
where $\Psi(f) \triangleq \mathcal{F}[s^{2}]$ and $\mathcal{F}$ denotes a Fourier transform and $s$ denotes the received signal \textit{after} filtering out-of-band distortion. If $s$ is also mixed to baseband using the nominal expected frequency $\tilde{f}$, the estimate $\hat{f}$ given in (\ref{eq:CFOest}) will correspond to the residual offset due to Doppler shift and mismatch between local oscillators in the transmitter and receiver $(f - \tilde{f})$. Ideally this value should be 0, indicating effective synchronization. 

By comparing the ratio of the desired CFO estimate peak magnitude to the highest-magnitude peak in any other frequency we can determine the robustness of synchronization for a given noise or interference power. We define this ``CFO Ratio'' as:
\begin{equation}\label{eq:CFOratiodefn}
    \mathrm{CFO\; Ratio\; [dB]} \triangleq 10 \log_{10} \left( \frac{ \Psi(\tilde{f}\,) } {{ \max_{f \neq \tilde{f}}} \Psi(f)}\right)
\end{equation}
Any value $>0$ dB would imply successful synchronization by identifying the appropriate frequency as the correct peak.

\section{Motivation}\label{sec:motivation}
\subsection{Dithering}
To illustrate the advantage of dithering, consider a sinusoidal signal with weak additive white Gaussian noise quantized to 3-bit resolution.
\begin{figure}[h]
\centering
\includegraphics[width=0.48\textwidth]{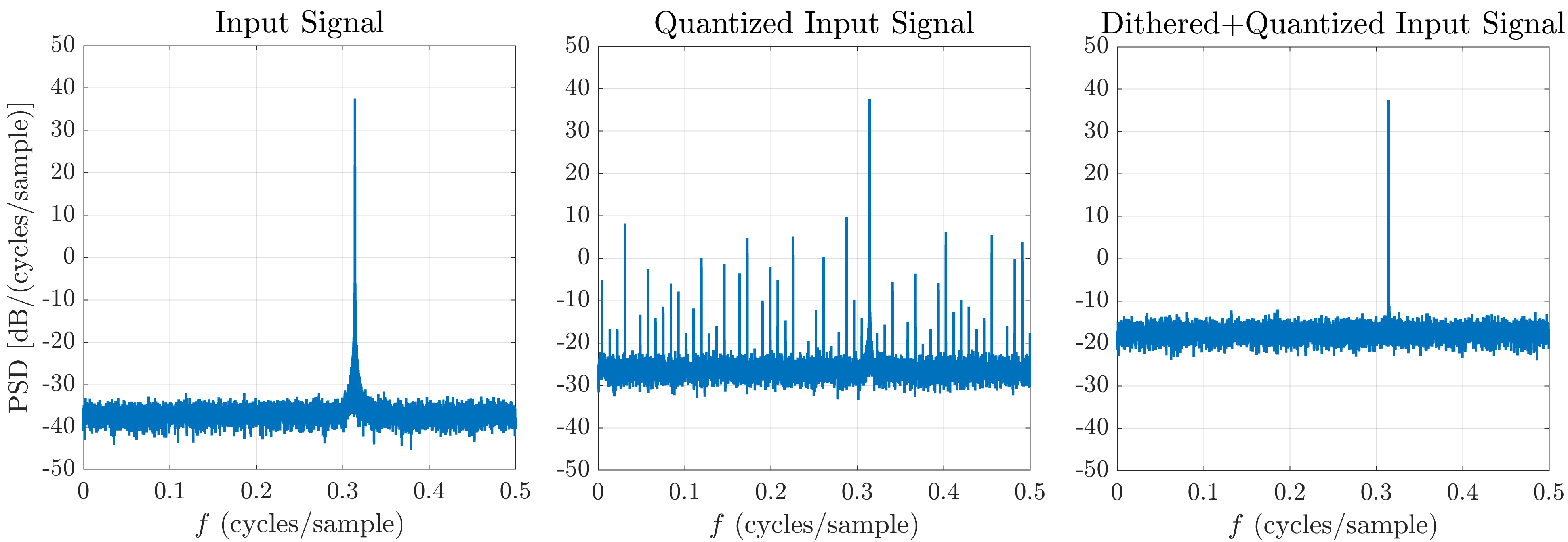}
\caption{Example PSD for Noisy Sinusoidal Input (Left) Quantized to 3-bit Resolution (Middle) and Quantized with Rectangular Dithering (Right)}
\label{fig:ditherpsdexample}
\end{figure}
Fig.~\ref{fig:ditherpsdexample} shows that without dithering the highly-correlated quantization error resulting from the low-resolution quantization generates harmonic ``spurs'' which alias throughout the output spectrum. 
By contrast, when using a uniform rectangular dither prior to quantization the output can maintain the same 3-bit resolution but with a significantly flatter spectral response, owing to the uncorrelated (spectrally white) quantization error. Consequently the SFDR of the system is improved despite an increase in the MSE (shown by the raised noise floor compared to the undithered quantization).
Formally, the process of rectangular dithering raises the MSE of the resultant quantized signal by 3 dB relative to the undithered quantization (from $\Delta^{2}/12$ to $\Delta^{2}/6$). But as shown by Fig.~\ref{fig:ditherpsdexample}, this 3 dB MSE compromise can support a 20+ dB improvement in SFDR. Since SFDR is only defined for sinusoidal inputs we evaluate all results on tone-like inputs, but the underlying advantage of dithering (uncorrelated quantization error) is widely applicable to many input signal types not studied here.

\subsection{Look-Up Table Architecture}\label{sec:LUTarch}
Here we propose an architecture to emulate the effects of dithering in the post-quantization digital-domain, thereby reducing the high cost and complexity required to implement analog-domain dithering.

\begin{figure}[h]
    \centering
    \includegraphics[width=0.46\textwidth]{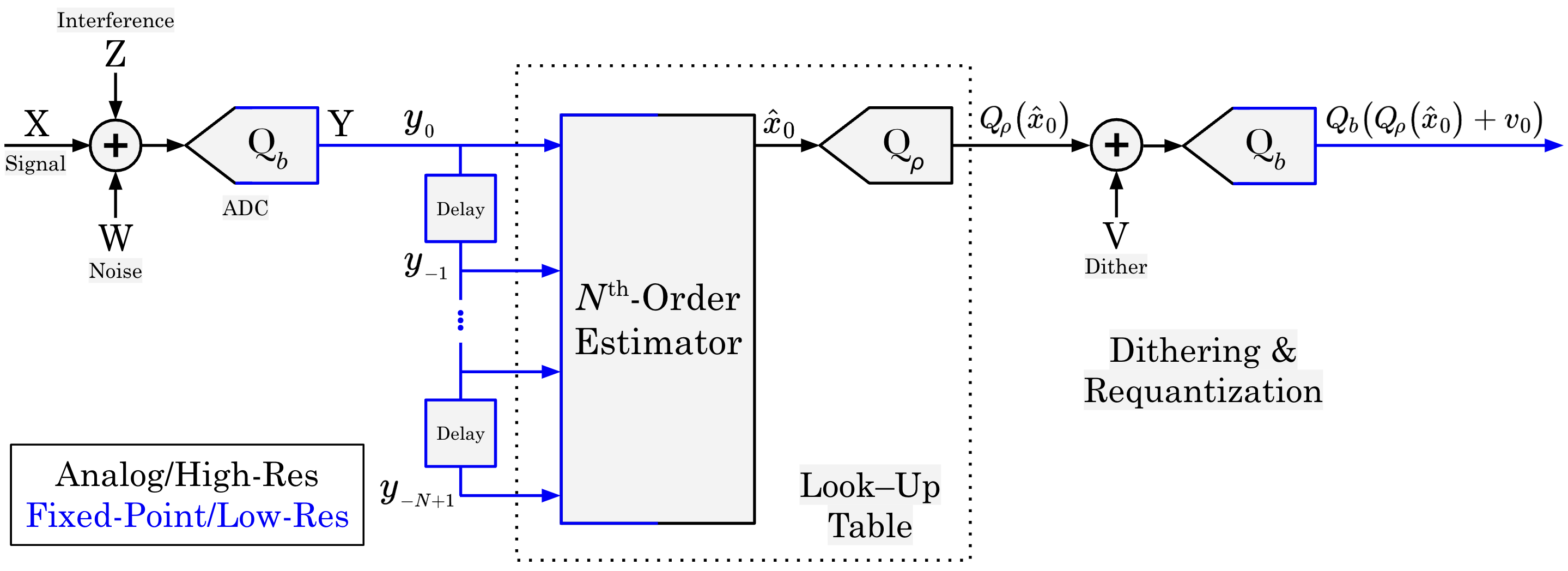}
    \caption{Implementation of Post-Quantization Dithering via Look-Up Table}
    \label{fig:ditherLUTBD}
\end{figure}

By using parametrized LUTs we can imitate the idea of 
dithering in the digital domain while emulating many of its desirable properties. 
LUT entries can be pre-computed to store an estimate $\hat{x}_{0}$ of the current analog input sample indexed by the $N$ previous quantized digital outputs $y_{n}, n \in \{-N+1, \cdots, 0\}$. The accuracy of this estimate can be improved by incorporating prior information about the input signal such as a parametric model and a prior distribution for its parameters.
The resultant high-resolution estimate can then be dithered according to the optimal rectangular dither distribution $p_{V}(v) = \mathsf{\Pi}_{\Delta}(v)$ and re-quantized to maintain the same fixed-point output precision as the original quantizer. 
Critically, the computationally-intensive estimation step can be encoded directly as a LUT entry which can be efficiently indexed for real-time application. Moreover, the resource-intensive dithering operation is thus implemented entirely in the digital-domain. 
A block diagram for the proposed scheme is given in Fig.~\ref{fig:ditherLUTBD}.

While this method is promising at a system-level, its performance is fundamentally limited by the estimation stage. This motivates the design and evaluation of an optimal estimator to be incorporated into this LUT-based correction. Results in this work will therefore inform the best possible performance of this LUT architecture, since the later low-resolution dithering and quantization stages will strictly increase the total error relative to the high-resolution estimator output.



\section{Analysis}\label{sec:analysis}

\subsection{Optimal Estimation}\label{sec:optimalest}
We model at time-index $n$ the instantaneous input 
to the quantizer as a sum of three independent sources. A desired signal $x_{n}$ (parametrized by $K$ parameters $\boldsymbol{\kappa}$), an interfering signal $z_{n}$ (parametrized by $U$ parameters $\boldsymbol{\mu}$), and additive white noise (or dither) $w_{n}$ (which we assume to be iid drawn from stationary distribution $p_{W}(w)$). Parameters $\boldsymbol{\kappa}$ and $\boldsymbol{\mu}$ are, in general, random variables. We denote the quantizer output: 
\begin{equation}
    y_{n} = Q_{b}(x_{n}(\boldsymbol{\kappa}) + z_{n}(\boldsymbol{\mu}) + w_{n})
\end{equation}

Define the 
set $\mathcal{I}_{b} \triangleq \left\{ 1, 2, \cdots, 2^{b} \right\}$ containing all possible quantization output indices. For analysis we assume without loss of generality that 
$C_{k} = k$ (which can later be isomorphically transformed in the digital-domain to arbitrary $C_{k}$, hence no loss of generality).

For an $N$\=/dimensional Look-Up Table we index each correction value entry using the previous $N$ output values. We therefore define the indexing vector $\mathbf{y} \in \mathcal{I}_{b}^{N}$ containing these samples as $\mathbf{y} = [y_{-N+1}, \cdots, y_{0}]^{T}$ such that $y_{[i]} = y_{-N+i}$.

When estimating the current analog input value $x_{0}$ using the observed output vector $\mathbf{y}$ we can use either minimum mean square error (MMSE) estimation, maximum-likelihood estimation (ML), or maximum-a-posteriori estimation (MAP). This gives rise to three formulations (as given in \cite{kay1993fundamentals}) for our estimation task:
\begin{align}
    \hat{x}_{0, \text{MMSE}}(\mathbf{y}) &= 
    \frac{\int_{-\infty}^{\infty}x_{0} \cdot p(x_{0}) \cdot p(\mathbf{y} | x_{0}) dx_{0}}{\int_{-\infty}^{\infty} p(x_{0}) \cdot p(\mathbf{y} | x_{0}) dx_{0}}\label{eq:mmseformulation}\\
    \hat{x}_{0, \text{ML}}(\mathbf{y}) &= \arg \max_{x_{0}} p(\mathbf{y} | x_{0})\label{eq:mlformulation}\\
    \hat{x}_{0, \text{MAP}}(\mathbf{y}) &= \arg \max_{x_{0}} 
    p(\mathbf{y} | x_{0}) \cdot p(x_{0})\label{eq:mapformulation}
\end{align}


Here $p(x_{0})$ is the prior distribution of the discretized (sampled) input signal at time $n=0$.
This function will, in general, depend on $\boldsymbol{\kappa}$ as given by (\ref{eq:priorx0kappa}). 



Notably, all of the expressions in (\ref{eq:mmseformulation}), (\ref{eq:mlformulation}), and (\ref{eq:mapformulation})
rely solely on the conditional distribution $p(\mathbf{y} | x_{0})$ and the prior distribution $p(x_{0})$, the expressions for which will depend on the parametric model used for the quantizer input signal. Here we also note that dependence on the prior distribution of $x_{0}$ makes these Bayesian estimators, including the conventionally frequentist ML estimator which relies on the prior distribution of parameters $p(\boldsymbol{\kappa})$ as will be shown next.

We can write the general expression for the conditional distribution as (see Appendix~\ref{sec:pyvcondx0derivation}):
\begin{multline}\label{eq:pyvcondx0_final}
    p(\mathbf{y} | x_{0}) = \idotsint_{\mathbb{R}^{U}} p(\boldsymbol{\mu}) 
    \cdot \idotsint_{\mathbb{R}^{K}} p(\boldsymbol{\kappa} | x_{0})\\
    \cdot \left(\prod_{n=-N+1}^{0} p(y_{n} | \boldsymbol{\kappa}, \boldsymbol{\mu})\right) d\boldsymbol{\kappa}\,d\boldsymbol{\mu} 
\end{multline}

with:
\begin{equation}\label{eq:pyncondku}
    p(y_{n} | \boldsymbol{\kappa}, \boldsymbol{\mu}) = \int_{T_{y_{n}} - x_{n}(\boldsymbol{\kappa}) - z_{n}(\boldsymbol{\mu})}^{T_{y_{n}+1} - x_{n}(\boldsymbol{\kappa}) - z_{n}(\boldsymbol{\mu})} p(w) dw 
\end{equation}

Likewise the prior distribution of the desired signal can be expressed intuitively in terms of its parameters:
\begin{equation}\label{eq:priorx0kappa}
    p(x_{0}) = \idotsint_{\mathbb{R}^{K}} p(\boldsymbol{\kappa})\cdot p(x_{0} | \boldsymbol{\kappa}) \,d\boldsymbol{\kappa} 
\end{equation}


Note that for Gaussian noise/dither $W \sim \mathcal{N}(0, \sigma^{2})$ we have:
\begin{equation}
    \int_{a}^{b} p(w) dw = \frac{1}{2}\left[\text{erf}\left(\frac{b}{\sigma \sqrt{2}}\right) - \text{erf}\left(\frac{a}{\sigma \sqrt{2}}\right) \right]
\end{equation}
for arbitrary $a,b$.


Note that for independently distributed parameters $\boldsymbol{\mu}$ we have $p(\boldsymbol{\mu}) = \prod_{i=1}^{U} p(\mu_{i})$ and in general $p(\boldsymbol{\kappa} | x_{0}) = \prod_{j=1}^{K} p(\kappa_{j} | \kappa_{j-1}, \cdots, \kappa_{1}, x_{0})$.

\subsection{Linear Estimation}
The linear minimum-mean square error estimator is given directly in closed-form (by \cite{kay1993fundamentals}) as:
\begin{equation}
    \hat{x}_{0,\mathrm{LMMSE}}(\mathbf{y}) = \mathbf{m}_{X_{0}\mathbf{Y}}^{T}\mathbf{M}^{-1}_{\mathbf{Y}\mathbf{Y}}(\mathbf{y} - \mathbb{E}[\mathbf{Y}]) + \mathbb{E}[X_{0}]
\end{equation}
where: 
\begin{align}
    \left(\mathbf{m}_{X_{0}\mathbf{Y}}\right)_{[i]} &= \mathbb{E}[(X_{0}-\mathbb{E}[X_{0}])(Y_{-N+i}-\mathbb{E}[Y_{-N+i}])]\\
    \left(\mathbf{M}_{\mathbf{Y}\mathbf{Y}}\right)_{[i,j]} &= \nonumber\\
    &\hspace{-7mm}\mathbb{E}[(Y_{-N+i}-\mathbb{E}[Y_{-N+i}])(Y_{-N+j}-\mathbb{E}[Y_{-N+j}])]
\end{align}
Each term in these expressions can be analytically pre-computed using (\ref{eq:pyvcondx0_final}) as: 
\begin{align}
    \mathbb{E}[X_{0}Y_{n}] &= \sum_{k=1}^{2^{b}} C_{k} \cdot \int_{\mathbb{R}} x_{0} \cdot p(x_{0}) \cdot p(y_{n} = C_{k} | x_{0}) dx_{0} \\
    \mathbb{E}[Y_{0}Y_{n}] &= \sum_{j=1}^{2^{b}} \sum_{k=1}^{2^{b}} C_{j} \cdot C_{k} \nonumber\\
    &\hspace{-3mm}\cdot \int_{\mathbb{R}} p(x_{0}) \cdot p(y_{0}=C_{j}|x_{0}) \cdot p(y_{n}=C_{k}|x_{0}) dx_{0}
\end{align}

The advantage of this linear estimator is that it can be implemented directly as a Finite Impulse Response (FIR) filter in the digital domain, consisting only of $N$ parallel multiplications of the input with filter coefficients followed by a summation. The disadvantage is that an FIR filter is impractical to implement for wideband devices as the computational overhead associated with multiplication is much higher than indexing a LUT and therefore infeasible at extremely high sampling rates. Nevertheless the LMMSE filter is included in some of our results as a baseline reference, representing the classic ``slow'' post-correction technique that can still be implemented as a sub-optimal LUT entry if so desired.

\subsection{Cramer-Rao Lower Bound}
The CRLB bounds the variance of the Minimum-Variance Unbiased Estimator \cite{kay1993fundamentals}. We use Bayesian estimators which can be biased, hence making Mean Square Error a more relevant criteria. The Bayesian Cramer-Rao Bound (BCRB) lower-bounds the MSE \cite{crafts_bayesian_2024} but is only a tight bound when the posterior distribution is Gaussian \cite{trees_detection_2013}, which our posterior distribution is not. No estimator will achieve a lower MSE than the MMSE estimator by definition, hence comparing it to any looser bound such as the BCRB is not informative and is therefore omitted.

\section{Methodology}\label{sec:methodology}
We constrain our evaluation to scenarios containing tone-like signals that are highly oversampled. This is intended to emulate the practical condition of a wideband receiver capturing some typical narrowband signals used in communication and radar. Namely, Secs.~\ref{sec:toneinput} and \ref{sec:toneinputtoneintf} address cases where the desired signal is a single-frequency tone, while Secs.~\ref{sec:bpskinput}, \ref{sec:bpskinputtoneintf}, and \ref{sec:bpskinputlfmintf} consider cases where the desired signal is a Binary-Phase Shift Keyed (BPSK) tone also with fixed frequency. The interference signal is modeled in Secs.~\ref{sec:toneinputtoneintf} and \ref{sec:bpskinputtoneintf} as a tone, while Sec.~\ref{sec:bpskinputlfmintf} models the interference as a Linear Frequency-Modulated (LFM) tone.
\begin{table}[h]
    \vspace{-4mm}
    \caption{Input Signal Models} \label{tbl:inputsignals}
    \vspace{-4mm}
    \begin{center}
        \begin{tabular}{ |c|c|c| }
            \hline
            \textbf{Desired} & \textbf{Interferer} & \textbf{Simulated Results}\\
            \hline
            Tone & - & Sec.~\ref{sec:toneinput}\\
            \hline
            BPSK & - & Sec.~\ref{sec:bpskinput}\\
            \hline 
            Tone & Tone & Sec.~\ref{sec:toneinputtoneintf}\\
            \hline
            BPSK & Tone & Sec.~\ref{sec:bpskinputtoneintf}\\
            \hline
            BPSK & LFM & Sec.~\ref{sec:bpskinputlfmintf}\\
            \hline
        \end{tabular}
    \end{center}
    \vspace{-4mm}
\end{table}
The set of desired and interference signal combinations in Table~\ref{tbl:inputsignals} is intentionally chosen to exercise several key properties of low-resolution quantization that complicate the estimation problem. These include in-band distortion of the signal of interest, in-band aliasing of harmonics from high-power out-of-band interference, and saturation (clipping) of the signal due to high total input power. The choice of LFM waveform also allows us to study time-variant quantization effects due to the sweeping interference frequency.

Next we give expressions and prior distributions for each of the signal models used as test inputs, which are to be substituted into the expressions in Sec.~\ref{sec:optimalest}. We further propose approximations for some of the intractable input signals which allow us to model them as a tone to more efficiently compute their estimate. In principle these approximations are just a numerical convenience, only provided to aid in reducing the time taken to pre-compute of estimates for every indexing sequence in the LUT. When a model used for tuning the estimator does not exactly match the actual input signal parameters we denote our approximated (model) value with a $\sim$ on top.
\subsection{Tone Input}\label{sec:toneinputmath}
\begin{align}
    x_{n,\mathrm{TONE}}(A, F, \Phi) &= A\cos(2\pi F n + \Phi) \nonumber\\
    z_{n,\mathrm{TONE}}(A_{\mathrm{z}}, F_{\mathrm{z}}, \Phi_{\mathrm{z}}) &= A_{\mathrm{z}}\cos(2\pi F_{\mathrm{z}} n + \Phi_{\mathrm{z}}) \nonumber
\end{align}

We assume uniform priors over scalar intervals $A \in [A_{\mathrm{L}}, A_{\mathrm{H}}]$, $F \in [F_{\mathrm{L}}, F_{\mathrm{H}}] \subseteq [0, 0.5]$, and $\Phi \in [0, 2\pi]$.

\begin{align}
    p(x_{0}) 
    &= \frac{1}{A_{\mathrm{H}} - A_{\mathrm{L}}} \int_{A_{\mathrm{L}}}^{A_{\mathrm{H}}} \frac{2a}{\pi \sqrt{a^{2} - {x_{0}}^{2}}} \cdot \mathsf{\Pi}_{2a}(x_{0}) da\\
    p(a | x_{0}) &= \mathsf{\Pi}_{(A_{\mathrm{H}}-A_{\mathrm{L}})}\left(a - \frac{A_{\mathrm{L}}+A_{\mathrm{H}}}{2}\right) \label{eq:pacondx0}\\
    p(f | a, x_{0}) &= \mathsf{\Pi}_{(F_{\mathrm{H}}-F_{\mathrm{L}})}\left(f - \frac{F_{\mathrm{L}}+F_{\mathrm{H}}}{2}\right) \label{eq:pfcondax0}\\
    p(\phi | f, a, x_{0}) &= \frac{1}{2} \sum_{m=0}^{1} \delta \left(\phi + (-1)^{m} \arccos\left(\frac{x_{0}}{a}\right)\right)\nonumber\\ 
    &\hspace{4cm}\cdot 2a \cdot \mathsf{\Pi}_{2a}(x_{0})\label{eq:pphicondfax0}
\end{align}

We also use uniform priors over the interference parameters:
\begin{equation}
    p(\boldsymbol{\mu}) = 
    p(a_{\mathrm{z}}) \cdot p(f_{\mathrm{z}}) \cdot p(\phi_{\mathrm{z}})
\end{equation}
with
\begin{align}
    p(a_{\mathrm{z}}) &= \mathsf{\Pi}_{(A_{\mathrm{z,H}}-A_{\mathrm{z,L}})}\left(a_{\mathrm{z}} - \frac{A_{\mathrm{z,L}}+A_{\mathrm{z,H}}}{2}\right) \\
    p(f_{\mathrm{z}}) &= \mathsf{\Pi}_{(F_{\mathrm{z,H}}-F_{\mathrm{z,L}})}\left(f_{\mathrm{z}} - \frac{F_{\mathrm{z,L}}+F_{\mathrm{z,H}}}{2}\right) \\
    p(\phi_{\mathrm{z}}) &= \mathsf{\Pi}_{2\pi}\left(\phi_{\mathrm{z}}\right)
\end{align}

Since this set of input signals is the most analytically-tractable, it will be used as a reference to design efficient approximate estimators for the other tone-like inputs. Consequently the prior distributions shown here are also reused unless otherwise specified.

\subsection{BPSK Input}\label{sec:bpskinputmath}
\begin{equation}
    x_{n,\mathrm{BPSK}}(A, F, \boldsymbol{\Theta}) = A\cos(2\pi F n + \Theta_{n})
\end{equation}
where $\boldsymbol{\Theta}$ is a stochastic process defined in terms of a Bernoulli process $\boldsymbol{R}$ as:
\begin{equation}\label{eq:Thetandefn}
    \Theta_{n} = \Phi + \pi \cdot \left[ \left(R_{\lfloor (n + L)/\tau \rfloor} - R_{0}\right) \textup{mod } 2 \right]
\end{equation}
where $\tau$ is the integer number of samples per symbol (period of the square pulse) which is assumed to be known exactly, $L$ is a discrete random variable shifting the sequence in time, and $\Phi$ is a continuous random variable representing the instantaneous phase offset of the BPSK signal.

The prior and conditional distributions for this input model are given in Appendix~\ref{sec:BPSKconddistder}. In general, this conditional distribution is too unwieldy and inconvenient to apply numerically. Thus we consider two approximations and evaluate their relative performance:

\subsubsection*{Case I: $N \ll \tau$} 
\begin{equation}\label{eq:BPSKtoneapprox}
    \widetilde{\Theta}_{n} = \Phi \approx \Theta_{n}
\end{equation}
In this case the window is sufficiently small that the BPSK signal can be approximated as a tone. This approximation can be applied for arbitrary M-ary Phase-Shift Keying (MPSK) modulation scheme and relies on the conditional distributions computed in Sec.~\ref{sec:toneinputmath}.

This tone approximation is exact in any window \textit{not} containing a phase transition and is inaccurate in any window that does contain a phase transition. The probability that the tone assumption is true for a given window is equal to the average proportion of windows not containing a phase transition which for arbitrary PSK modulation order $M$ is given by (see Appendix~\ref{sec:toneassumptionprobderivation}): 
\begin{multline}\label{eq:probToneMPSK}
    p\left(\boldsymbol{\Theta} = \Phi\right)\\
    = \left(\frac{1}{M}\right)^{\left\lceil N/\tau \right\rceil} \cdot \left[1 + (M-1) \left(\left\lceil \frac{N}{\tau} \right\rceil - \frac{N-1}{\tau}\right) \right]
\end{multline}



When $N \leq \tau$ we can express (\ref{eq:probToneMPSK}) as:
\begin{equation}\label{eq:NlTprobvalidtoneMPSK}
    p\left(\boldsymbol{\Theta} = \Phi\right) = 1 - \frac{M-1}{M}\cdot \frac{N-1}{\tau}
\end{equation}

For BPSK we have $M=2$ thus (\ref{eq:NlTprobvalidtoneMPSK}) simplifies to:
\begin{equation}\label{eq:NlTprobvalidtoneBPSK}
    p\left(\boldsymbol{\Theta} = \Phi\right)
    = 1 - \frac{N - 1}{2\tau}
\end{equation} 

Using (\ref{eq:probToneMPSK}) we can evaluate the accuracy of this assumption for BPSK inputs as a function of the LUT window $N$ and oversampling rate $\tau$. 
For example, to maintain the tone assumption validity for at least 90\% of windows the period of the BPSK signal must be $\gtrsim 5 \times$ the window length.

\subsubsection*{Case II: $N \leq \tau$}
\begin{equation}\label{eq:BPSKexact}
    \widetilde{\Theta}_{n} = \Theta_{n}, n = \left\{-N+1, \cdots, 0\right\}
\end{equation}
In this case the BPSK signal is guaranteed to undergo at most one phase transition within the window length $N$ of the LUT. This significantly limits the ensemble of sequences $\boldsymbol{\Theta}$ and allows efficient direct computation of the conditional distribution, as given in Appendix~\ref{sec:BPSKexactconddist}.

\subsection{LFM Input}\label{sec:lfminputmath}
\begin{align}
    z_{n,\mathrm{LFM}}(A_{\mathrm{z}}, F_{\mathrm{z}}, \Phi_{\mathrm{z}}, \Omega_{\mathrm{z}}, \Upsilon_{\mathrm{z}}) &= \nonumber\\
    &\hspace{-36mm}A_{\mathrm{z}}\cos\biggl(2\pi \left(F_{\mathrm{z}} - \frac{\Omega_{\mathrm{z}}}{2} + \frac{\Omega_{\mathrm{z}}}{\Upsilon_{\mathrm{z}}} \left(n \text{ mod } \Upsilon_{\mathrm{z}}\right)\right) n 
    + \Phi_{\mathrm{z}}\biggr)
\end{align}

Due to the large number of parameters describing this model, we opt to apply the approximation:
\begin{equation}
    \left(F_{\mathrm{z}} - \frac{\Omega_{\mathrm{z}}}{2} + \frac{\Omega_{\mathrm{z}}}{\Upsilon_{\mathrm{z}}} \left(n \text{ mod } \Upsilon_{\mathrm{z}}\right)\right) \approx F_{\mathrm{z}}
\end{equation}
which is accurate for narrowband LFM signals (small $\Omega_{\mathrm{z}}$). Moreover, for narrow windows ($N \ll \Upsilon_{\mathrm{z}}$) we can confidently approximate $z_{n,\mathrm{LFM}}$ as a tone with unknown instantaneous frequency ($F \in [F_{\mathrm{z}} - \Omega_{\mathrm{z}}/2, F_{\mathrm{z}} + \Omega_{\mathrm{z}}/2]$). The maximum deviation from this assumption within a LUT window of size $N$ not overlapping with an LFM pulse repetition boundary is $\Omega_{\mathrm{z}} N / (2\Upsilon_{\mathrm{z}})$. The advantage of this approximation is that we can re-use the conditional and prior distributions from Sec.~\ref{sec:toneinputmath} (ignoring $\Omega_{\mathrm{z}}$ and $\Upsilon_{\mathrm{z}}$), only changing the prior frequency distribution to a uniform one within the deviation range.


\section{Simulated Results}
\subsection{Simulation Setup}\label{sec:simulationsetup}
In the following sections we will present simulated results showcasing the capability of our proposed estimators, each of which uses the estimation methods detailed in Sec.~\ref{sec:optimalest} applying the distributions defined in Sec.~\ref{sec:methodology}. 

Unless otherwise specified each simulation uses $10^{5}$ samples, uniform mid-riser quantization with $b=3$-bit resolution ($\Delta = 0.25$) denoted $Q(.)$, iid Gaussian input noise $w_{n} \sim \mathcal{N}(0, \sigma^{2})$, $\sigma/\Delta = 0.16$, $A = 1-\Delta/2 = 0.875$, $F = \pi/10$, and $\Phi = 0$. Digital dither samples are generated iid from a rectangular distribution $p(v) = \mathsf{\Pi}_{\Delta}(v)$. Frequencies for training and testing of the LUT are always non-integer submultiples of the sampling rate, ensuring ergodicity of the resultant signals. This is achieved by using irrational frequencies, usually done by incrementing a rational frequency by $\pi/1000$ or by using angular frequency $2\pi F = \pi^{2} / 5$ directly. All simulation results use known prior distributions with $A_{\mathrm{L}} = A_{\mathrm{H}} = A$, a reasonable assumption due to the prevalence of Automatic Gain Control (AGC) in real-world ADCs which ensures normalization of input amplitude to a known value. As a reference value we sometimes present results denoted $N=0$ for which no LUT is used, in which case $\hat{x}_{0} = y_{0}$.

Another note of relevance to the simulations is that computation of the ML estimate is much more efficient than the MMSE estimate, since the former can use numerical optimization techniques while the latter requires direct computation of a dense grid of $x_{0}$ values for accurate integration. For this reason results in the later sections are presented predominantly using the ML estimator due to high computational complexity of the associated estimation problems. Choice of estimator is always stated explicitly for each result.

\subsection{Tone Input}\label{sec:toneinput}

\begin{figure}[h]
    \centering
    \includegraphics[width=0.48\textwidth]{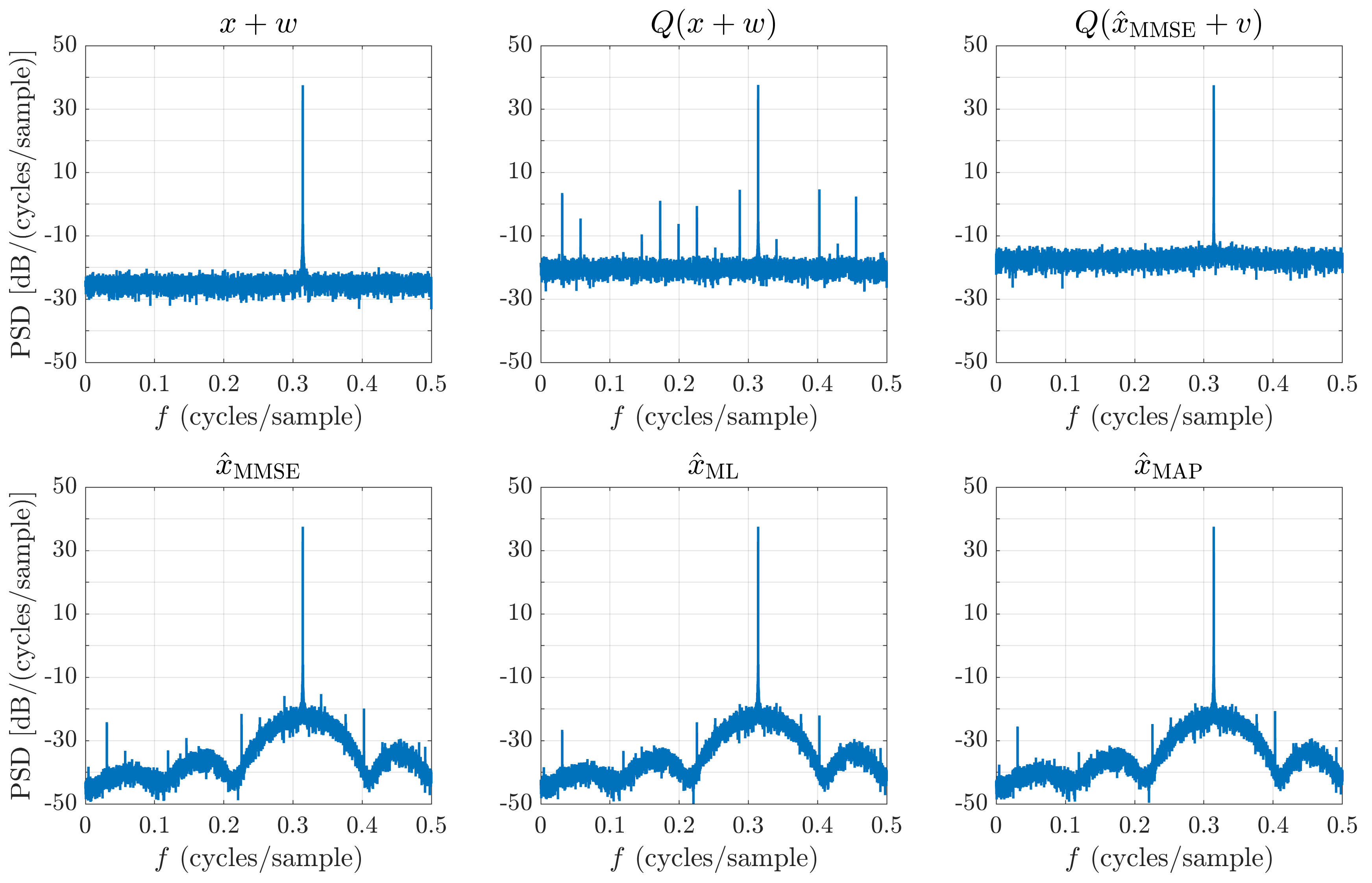}
    \caption{Comparison of Power Spectral Densities at Each LUT Stage ($N=10$)}
    \label{fig:label2}
\end{figure}

A LUT is trained according to the model outlined in Sec.~\ref{sec:toneinputmath}, and applied to a tone input with parameters described in Sec.~\ref{sec:simulationsetup}. The effect of the LUT is illustrated by the PSD at each stage in Fig.~\ref{fig:label2}, which compares the original noisy ADC input, the quantized ADC output, the output of all three estimators, and the output of the MMSE estimator after requantization with rectangular dithering. The results indicate a significant improvement in spectral purity after estimation which is preserved even at the fixed-point output of the LUT.

\begin{figure}[h]
    \centering
    \includegraphics[width=0.48\textwidth]{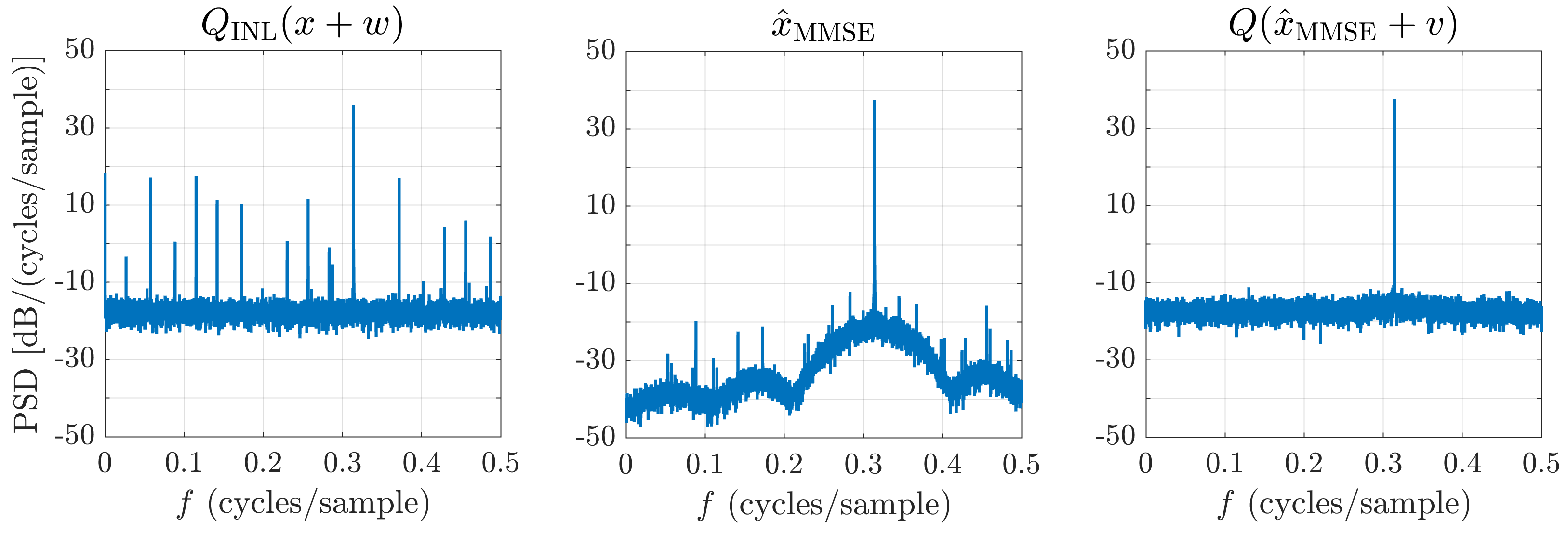}
    \caption{Comparison of Power Spectral Densities at Each LUT Stage with Non-Linear Quantization ($N=10$)}
    \label{fig:label2p5}
\end{figure}

By expressing each estimator in our generalized form accepting arbitrary $\mathbf{T}$, we can accommodate correction even of highly non-linear quantizers. 
As an example, consider the same quantizer used previously but corrupted by INL sequence: 
$\mathbf{INL} = [-0.4, -1.3, -0.7, 0.2, 1.1, 0.3, -0.3]\cdot \Delta$.
We denote this new non-linear transfer function as $Q_{\mathrm{INL}}(.)$. The resultant quantized signal power spectrum in Fig.~\ref{fig:label2p5} shows much stronger harmonic content distributed over more frequency components. Nevertheless, our estimator succeeds in reducing these below the noise floor at the requantized LUT output. This demonstrates the versatility of our estimator for handling arbitrary quantizers including those with highly non-ideal transfer functions.

\subsubsection*{Sensitivity Analysis}
\begin{figure}[h]
    \centering
    \includegraphics[width=0.48\textwidth]{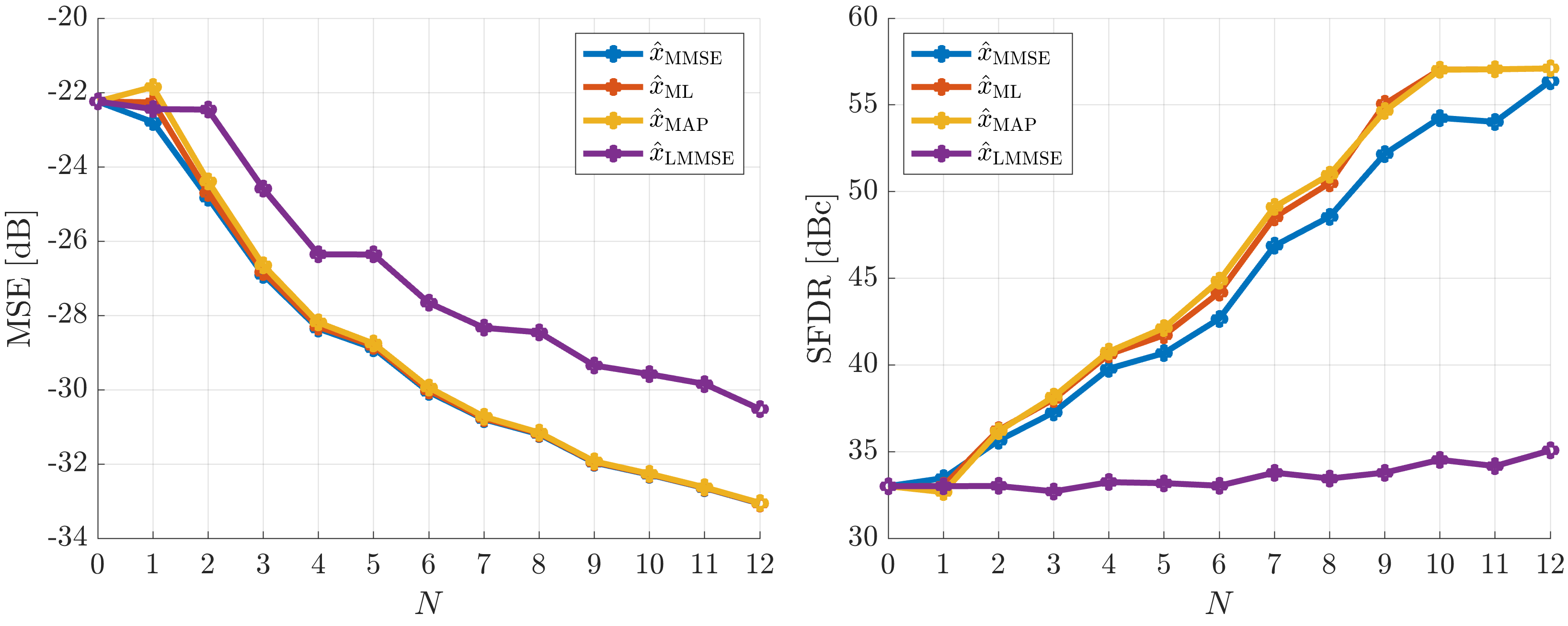}
    \caption{LUT Tone Estimator Performance over $N$ 
    }
    \label{fig:label3}
\end{figure}

Sensitivity of LUT performance to the value of $N$ on a tone input is quantified in Fig.~\ref{fig:label3}, demonstrating the consistent ability of all three non-linear estimators to reduce MSE by increasing $N$. Notably the SFDR increase levels off above $N \approx 9$ for this set of input parameters, as the spurs are entirely reduced below the noise floor and the SFDR improvement becomes limited. Moreover, the linear MMSE estimator is shown to be significantly less effective in MSE reduction (up to $3$ dB worse) and vastly less effective in SFDR improvement ($>15$ dB worse). Due to this clear inadequacy we omit LMMSE results from all later results. Also of note is the inconsistent rate of MSE improvement, indicating that some samples are more information-theoretically valuable for estimation than others. This observation forms the basis for bit-masking, a topic for future work.

\begin{figure}[h]
    \centering
    \includegraphics[width=0.48\textwidth]{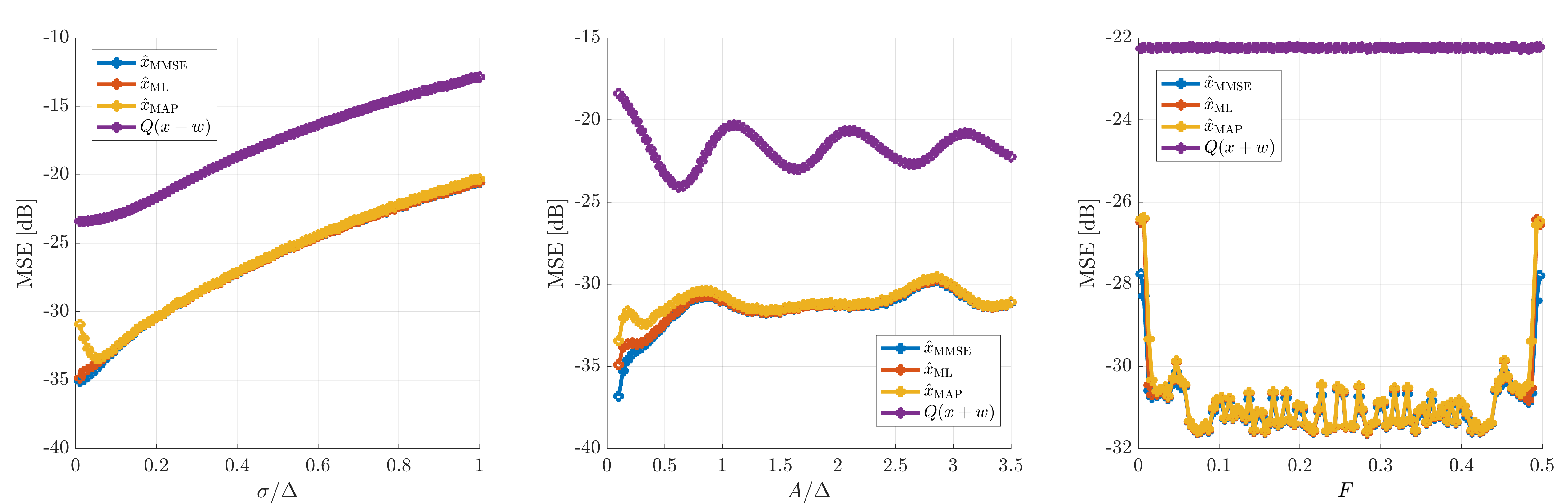}
    \caption{LUT Tone Estimator Performance over $A,\sigma,F$ with Values Known Exactly A-Priori ($N=8$)}
    \label{fig:label4}
\end{figure}

To further validate LUT performance, we plot sensitivity to input parameters for $\widetilde{A} = A, \widetilde{\sigma} = \sigma, \widetilde{F} = F$ when all three values are known a-priori exactly ($A_{\mathrm{L}} = A_{\mathrm{H}} = A, F_{\mathrm{L}}=F_{\mathrm{H}}=F$) in Fig.~\ref{fig:label4}. Notable takeaways from these plots are consistency of MSE improvement over $\sigma$ even for large noise values, and consistency of MSE improvement over $F$ except in cases when $N \ll 1/F$ due to slow variation relative to LUT window size.

\begin{figure}[h]
    \centering
    \includegraphics[width=0.48\textwidth]{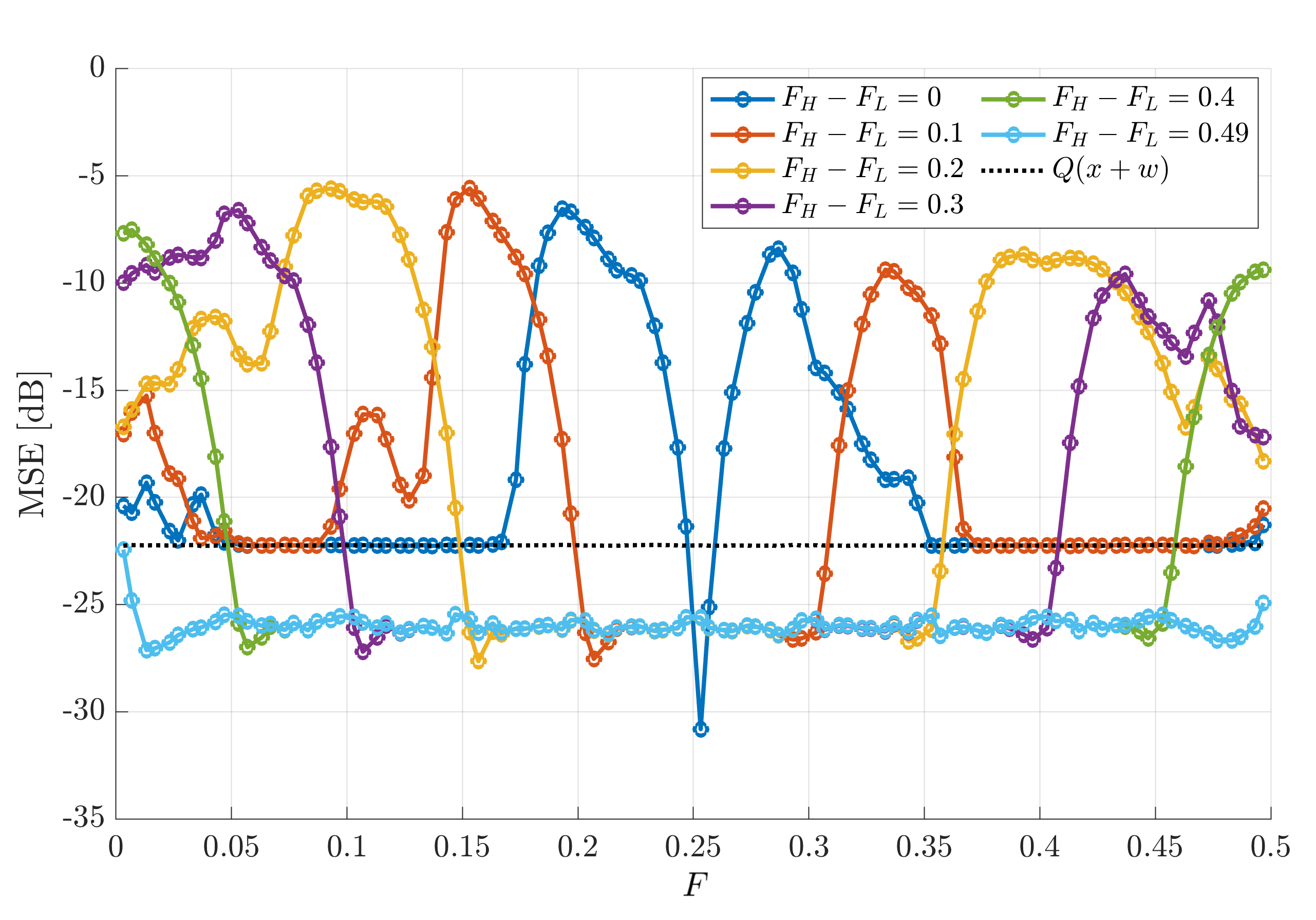}
    \caption{LUT Tone Estimator Performance over $F$ Using $\hat{x}_{\mathrm{MMSE}}$ with $\widetilde{F} = 1/4 + \pi/1000$, Varying Training Bandwidth with $F_{\mathrm{H}} - \widetilde{F} = \widetilde{F} - F_{\mathrm{L}}$, ($N=8$)}
    \label{fig:label5}
\end{figure}

Another topic of interest is sensitivity of the LUT to model mismatch, such as a table trained for nominal frequency $\widetilde{F}$ but tested with $F \neq \widetilde{F}$. This result is shown in Fig.~\ref{fig:label5}, indicating that by adjusting our prior distribution when training the LUT we can achieve a consistent improvement in MSE over a wide range of input frequencies at the expense of reduced improvement in any individual frequency of interest. Notably a single pre-trained LUT is shown to be capable of correcting arbitrary input frequency across the entire bandwidth of the receiver.


\subsection{BPSK Input}\label{sec:bpskinput}
\begin{figure}[h]
    \centering
    \includegraphics[width=0.48\textwidth]{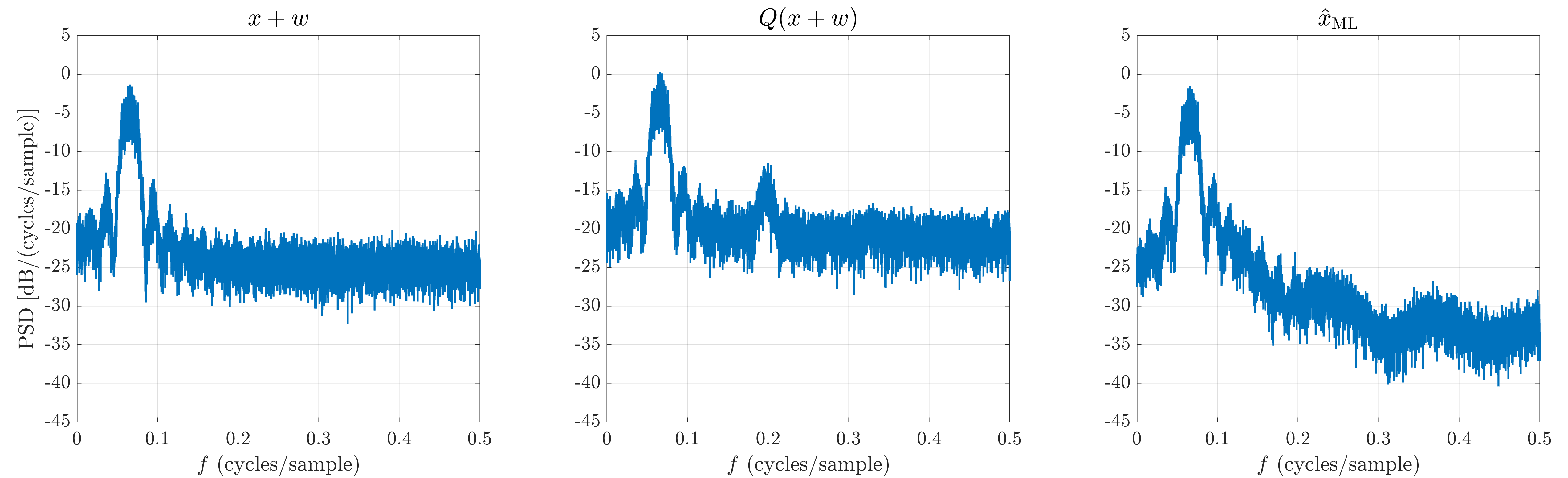}
    \caption{BPSK Estimator PSD Comparison ($A=0.125, F=1/16 + \pi/1000, \tau=50, N=8$)}
    \label{fig:label8}
\end{figure}

Next, we consider a communication signal consisting of a binary phase-shift keyed (BPSK) sinusoid with a square pulse shape.
We can evaluate the exact estimation method using the model in (\ref{eq:BPSKexact}) by computing its output PSD, shown by Fig.~\ref{fig:label8} which demonstrates excellent cancellation of the prominent third-harmonic spur.

\subsubsection*{Sensitivity Analysis}
\begin{figure}[h]
    \centering
    \includegraphics[width=0.48\textwidth]{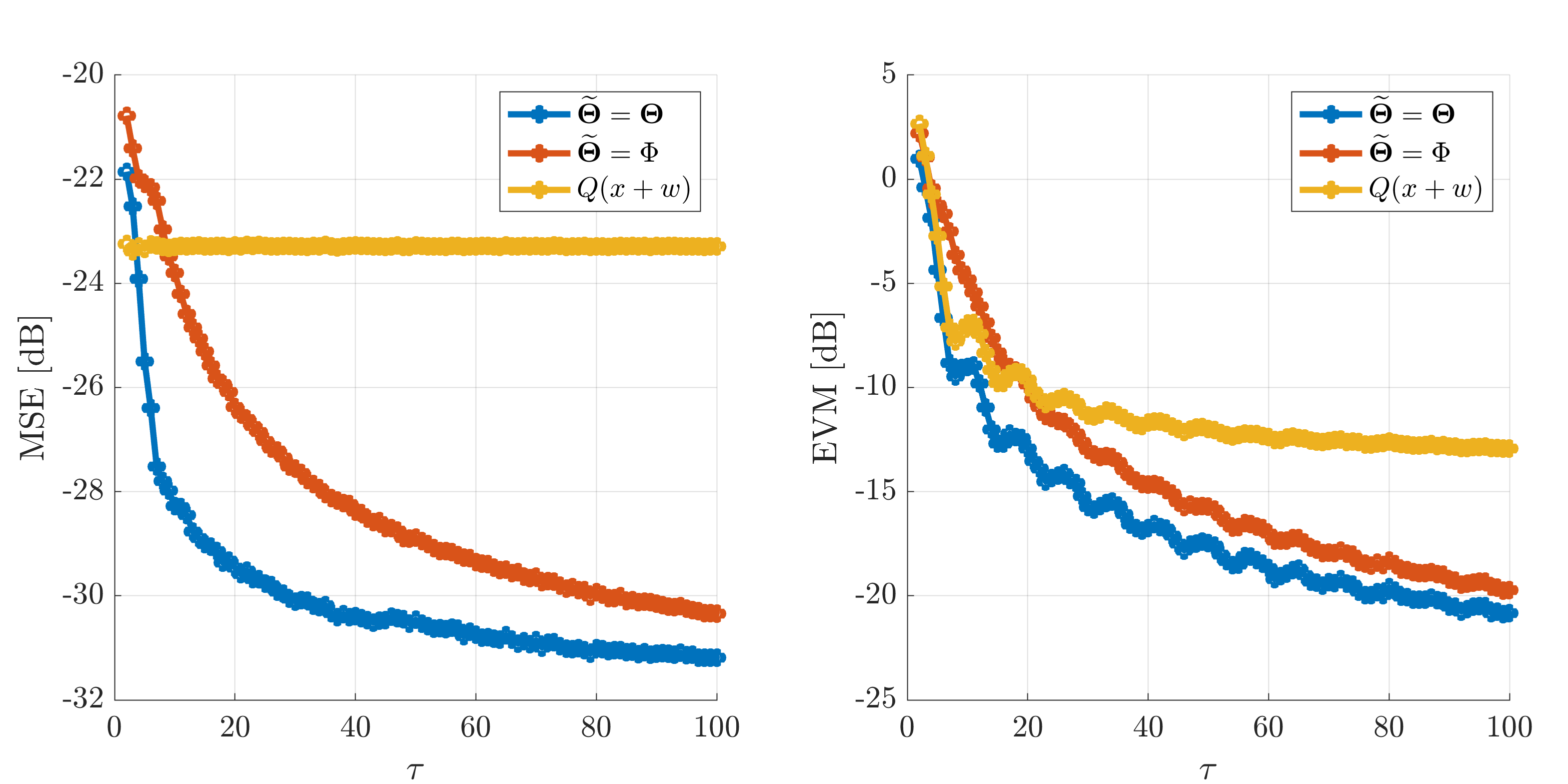}
    \caption{BPSK Estimator Performance over $\tau$ Using $\hat{x}_{\mathrm{ML}}$ ($A=0.125, F = 1/16 + \pi/1000, N=8$)}
    \label{fig:label9}
\end{figure}

\begin{figure}[h]
    \centering
    \includegraphics[width=0.48\textwidth]{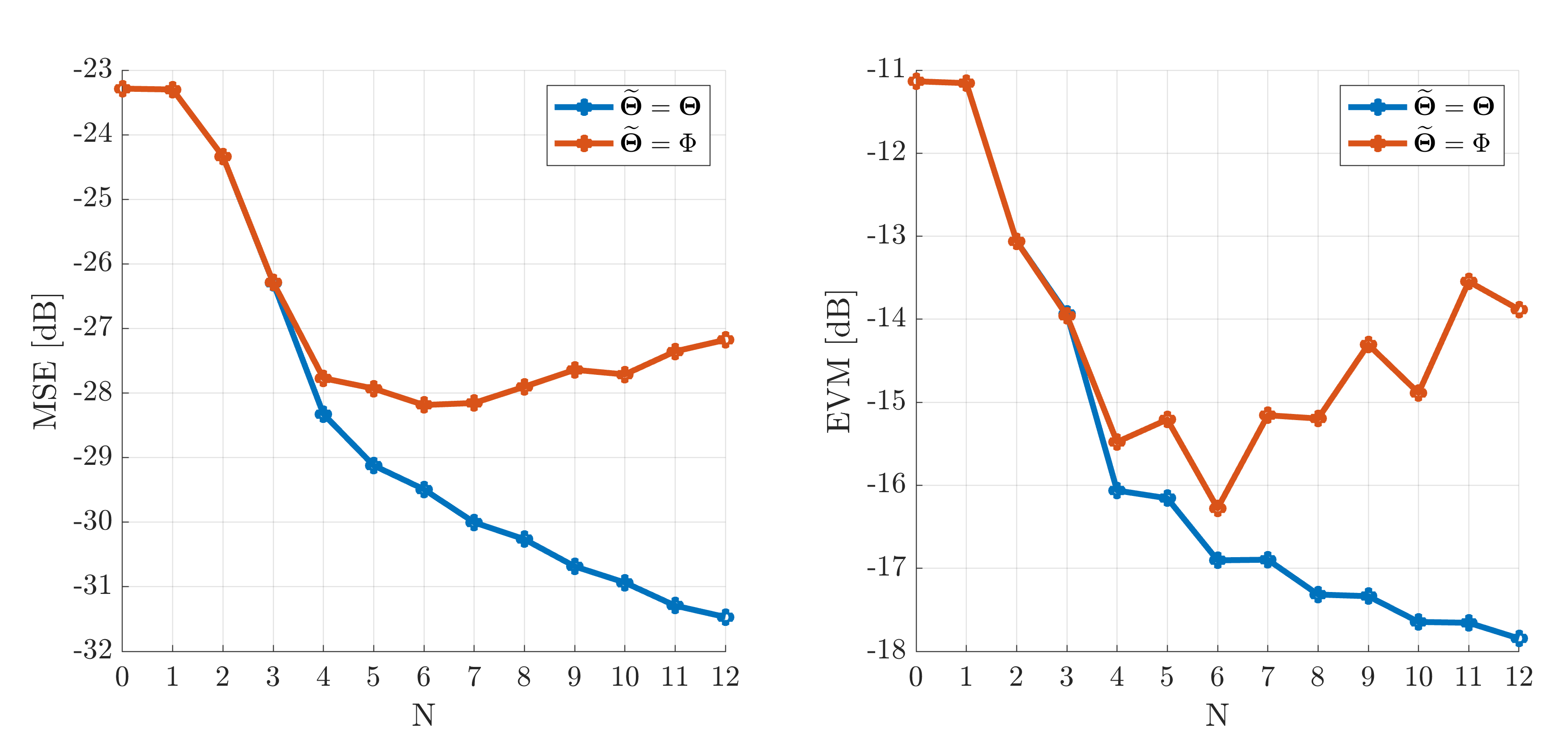}
    \caption{BPSK Estimator Performance over $N$ Using $\hat{x}_{\mathrm{ML}}$ ($A=0.125, F = 1/4 + \pi/1000, \tau = 50$)}
    \label{fig:label10}
\end{figure}

Sensitivity of both the exact estimator (\ref{eq:BPSKexact}) and the tone-approximation (\ref{eq:BPSKtoneapprox}) to $\tau$ is shown in Fig.~\ref{fig:label9}, showing a consistent relative improvement in both MSE and EVM by using the exact (optimal) estimator. This is further evidenced by Fig.~\ref{fig:label10} showing sensitivity of the same metrics to $N$. In the latter figure, one notable takeaway is that for large $N$ the tone assumption is increasingly unlikely to be satisfied hence the worsening performance above $N \approx 6$ when not using the exact method. The trade-off is that the optimal estimator requires increased computational complexity to evaluate during pre-training.

Both estimators are most applicable to wideband receivers which have a high oversampling rate relative to the bandwidth of any BPSK signal of interest (large $\tau$). This allows the use of relatively long windows (large $N$) while maintaining the assumption $N \leq \tau$ required for accurate estimation using the exact method in (\ref{eq:BPSKexact}).






\subsection{Tone Input with Tone Interferer}\label{sec:toneinputtoneintf}
\begin{figure}[h]
    \centering
    \includegraphics[width=0.48\textwidth]{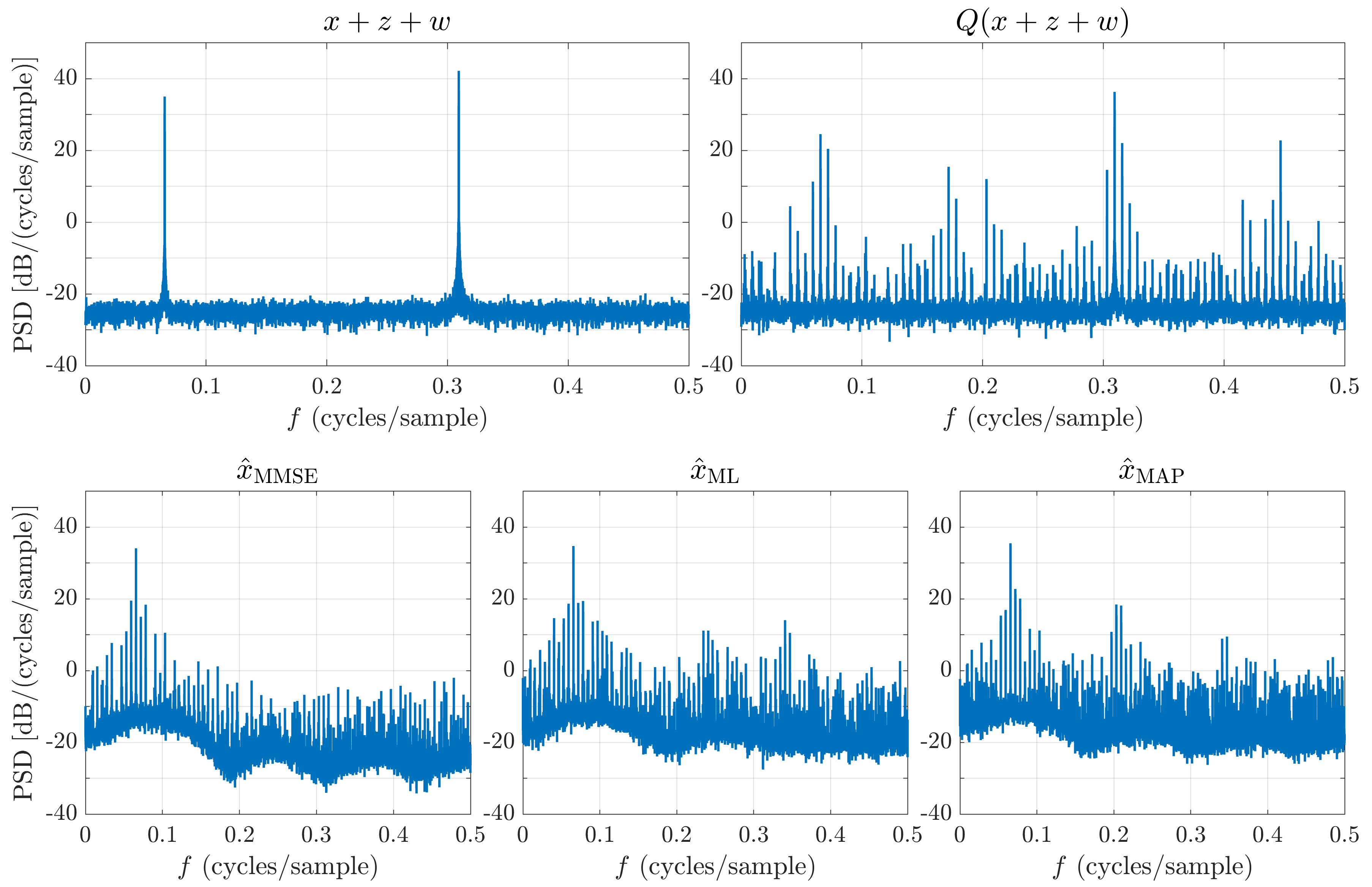}
    \caption{Tone Input with Tone Interferer PSD Comparison ($A=0.875, F = 1/16 + \pi/1000, A_{\mathrm{z}} = 2, F_{\mathrm{z}} = 5/16 - \pi/1000, N=8$)}
    \label{fig:label11}
\end{figure}

In this section we consider a desired tone input with a simultaneous tone interferer elsewhere in the input spectrum.
The interference case is particularly relevant to wideband communication devices, in which the prevalence of multiple high-power signal sources will frequently saturate the ADC and result in high-power harmonic and intermodulation products. This is illustrated in Fig.~\ref{fig:label11}, which demonstrates three key properties of the proposed estimators. First, successful cancellation of the unwanted interfering signal. Second, successful reduction in the power of harmonic and intermodulation spurs resulting from low-resolution saturated quantization. Third, restoration of power in the desired signal which is lost by the saturation process. 


\subsubsection*{Sensitivity Analysis}
\begin{figure}[h]
    \centering
    \includegraphics[width=0.48\textwidth]{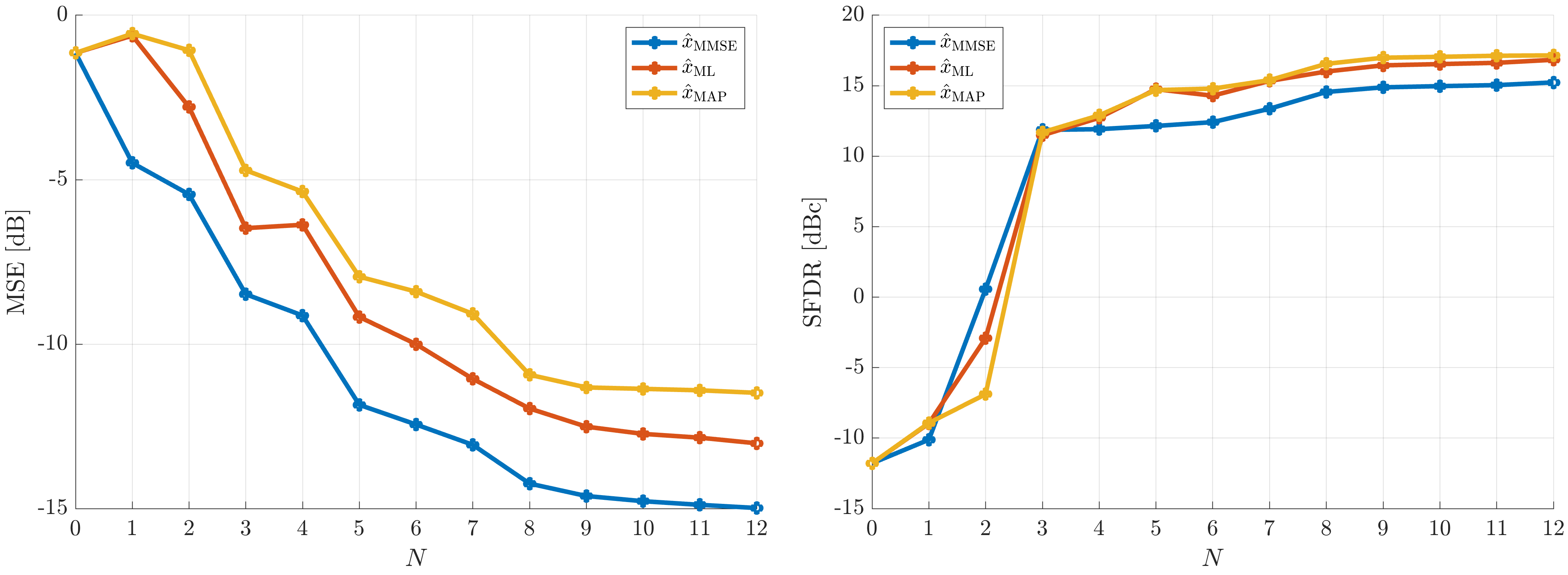}
    \caption{Tone+Tone Estimator Performance over $N$ ($A=0.875, F = 1/16 + \pi/1000, A_{\mathrm{z}} = 2, F_{\mathrm{z}} = 5/16 - \pi/1000$)}
    \label{fig:label12}
\end{figure}

The sensitivity of the proposed estimator to $N$ is plotted in Fig.~\ref{fig:label12}, indicating that despite the several dB improvement in MSE offered by the MMSE estimator, it consistently equals or underperforms the ML/MAP estimators in SFDR improvement especially for $N \geq 5$. 

\begin{figure}[h]
    \centering
    \includegraphics[width=0.48\textwidth]{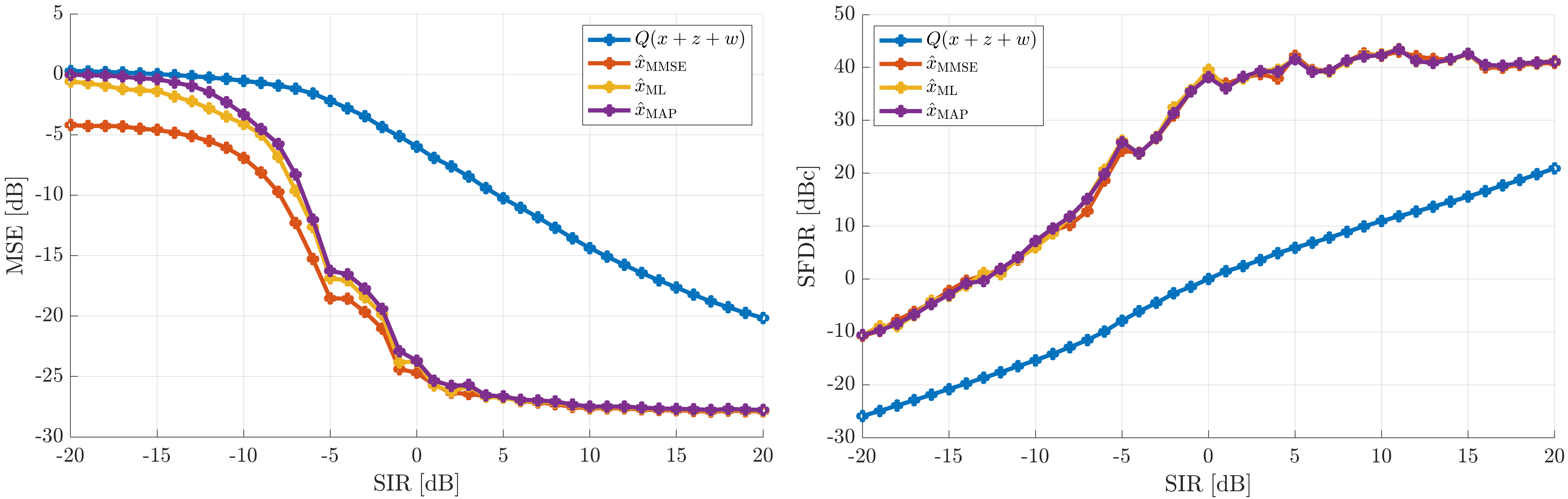}
    \caption{Tone+Tone Estimator Performance over Signal-to-Interference Power Ratio ($A=0.875, F = 1/16 + \pi/1000, F_{\mathrm{z}} = 5/16 - \pi/1000, N=5$)}
    \label{fig:label13}
\end{figure}

We define the input Signal-to-Interference Ratio (SIR) as $\mathrm{SIR\; [dB]} \triangleq 10 \log_{10} (\mathbb{E}[X^{2}]/\mathbb{E}[Z^{2}])$ and show its effect on estimator performance in Fig.~\ref{fig:label13}. As expected, it is easier to cancel low-power interference due to minimal saturation. However even with high-power interference where little MSE gain is possible there is still a non-negligible SFDR improvement in the resultant PSD.

\subsection{BPSK Input with Tone Interferer}\label{sec:bpskinputtoneintf} 

\begin{figure}[h]
    \centering
    \includegraphics[width=0.48\textwidth]{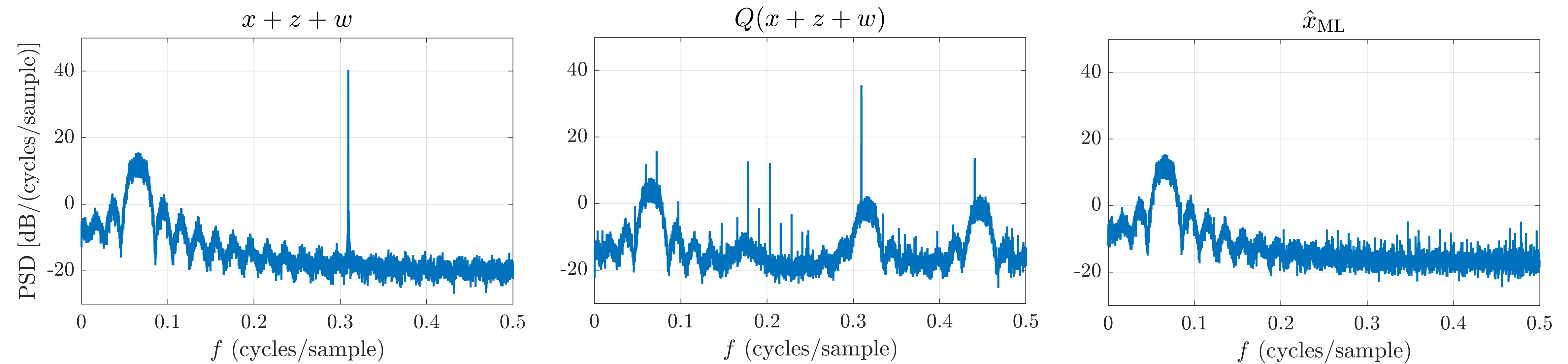}
    \caption{BPSK+Tone Estimator PSD Comparison ($A=0.875, F = 1/16 + \pi/1000, \tau=50, A_{\mathrm{z}} = 1.6, F_{\mathrm{z}} = 5/16 - \pi/1000, N=12$)}
    \label{fig:label14}
\end{figure}

The presence of high-power narrowband interference with low-power wideband inputs is a particularly difficult spectral scenario for conventional post-processing algorithms to correct. In the case illustrated by the PSDs in Fig.~\ref{fig:label14}, the third harmonic of the high-power (saturating) tone interferer aliases into the bandwidth of the desired BPSK signal. Consequently, no amount of linear filtering to remove out-of-band distortion will compensate for this in-band harmonic. Moreover, the BPSK power is significantly reduced to saturation of the ADC and the remaining spectrum is heavily distorted by harmonic and intermodulation products. Nevertheless, a 12th-order ML estimator succeeds in addressing all of these issues, recovering an excellent estimate of the original BPSK signal. 

\begin{figure}[h]
    \centering
    \includegraphics[width=0.48\textwidth]{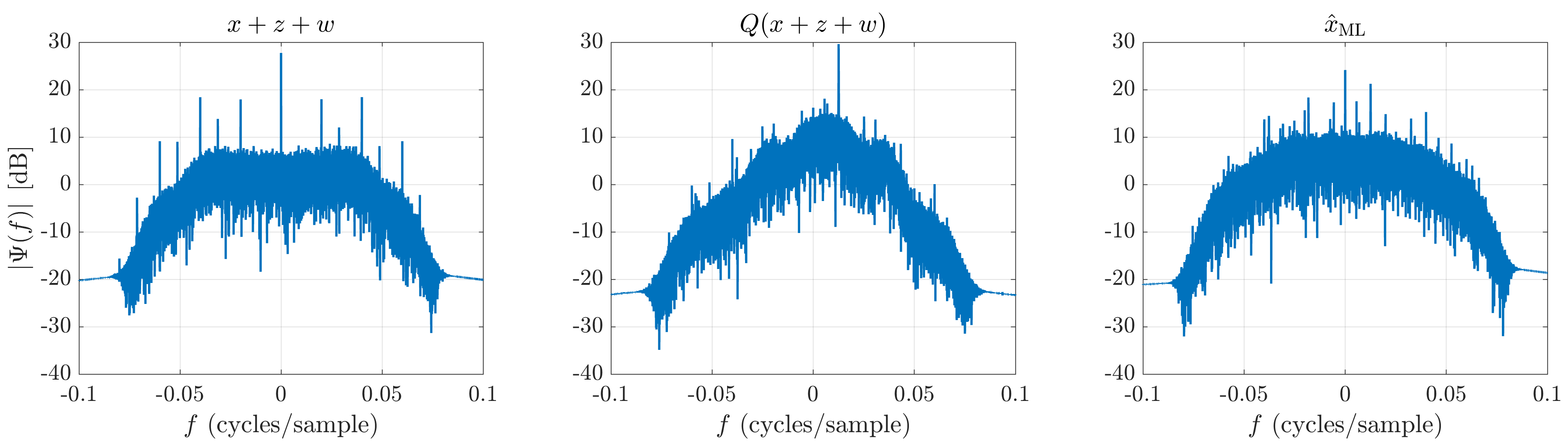}
    \caption{Carrier Frequency Offset Estimation for BPSK+Tone ($A=0.16, F = 1/16 + \pi/1000, \tau=50, A_{\mathrm{z}} = 1.6, F_{\mathrm{z}} = 5/16 - \pi/1000, N=12$)}
    \label{fig:label15}
\end{figure}

The presence of in-band distortion due to higher-order interference harmonics aliasing into our BPSK signal bandwidth disrupts this synchronization process, as illustrated in Fig.~\ref{fig:label15} where the quantized ADC output has a maximum that is not located at $f=0$. After applying our ML estimator the maximum peak is at the appropriate value of $0$, indicating a sucessful synchronization made possible by our post-processing method.

\begin{figure}[h]
    \centering
    \includegraphics[width=0.48\textwidth]{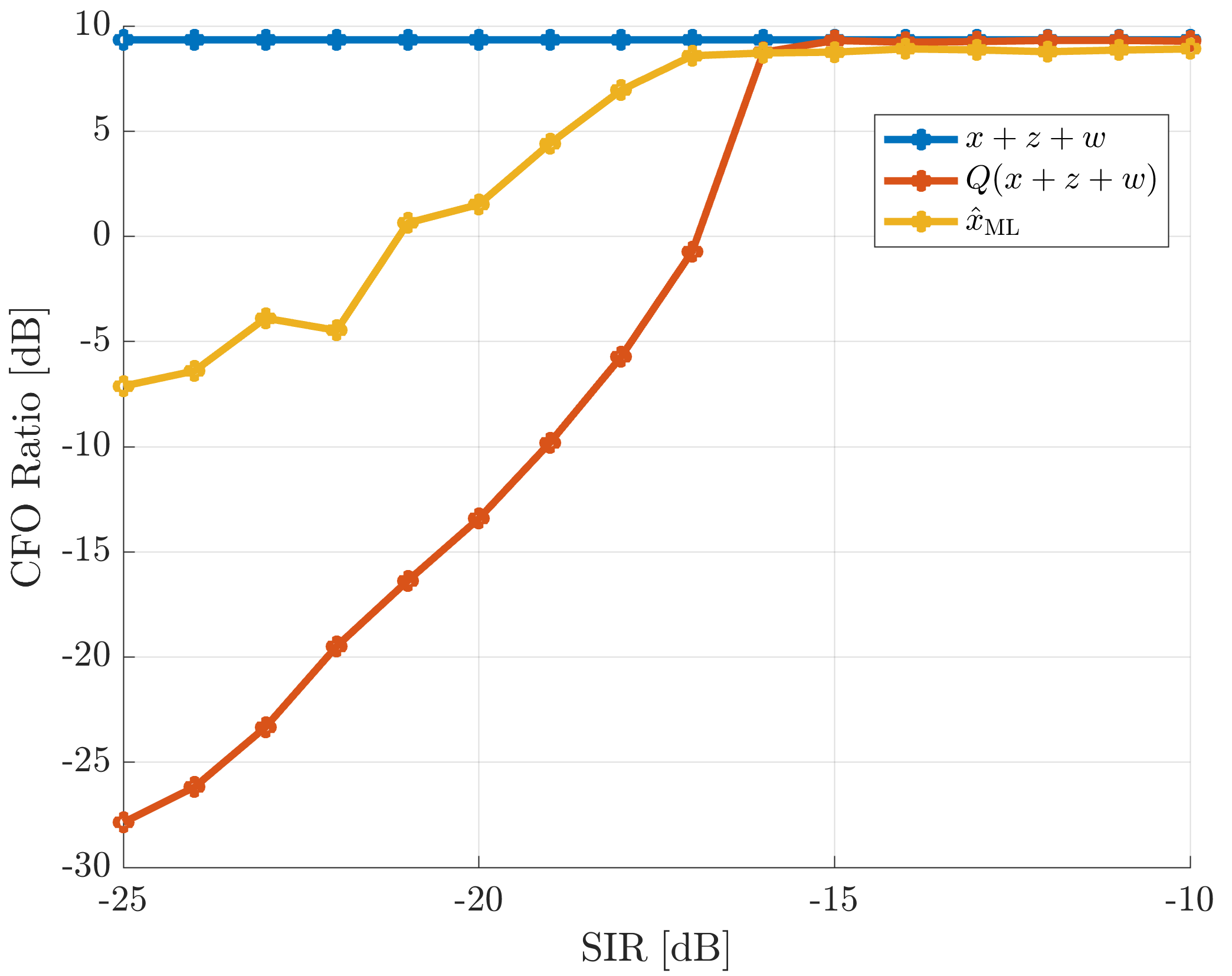}
    \caption{BPSK+Tone Estimator Performance over Signal/Interference Power Ratio ($A=0.16, F = 1/16 + \pi/1000, \tau=50, F_{\mathrm{z}} = 5/16 - \pi/1000, N=8$)}
    \label{fig:label16}
\end{figure}

The CFO Ratio (defined in (\ref{eq:CFOratiodefn})) is shown as a function of SIR in Fig.~\ref{fig:label16}. Synchronization is achieved when the CFO Ratio is $>0$ dB.
Notably the use of our post-processing estimation method in this particular spectral scenario allows synchronization with $4$ dB lower SIR when compared to the raw quantized output.

\subsection{BPSK Input with LFM Interferer}\label{sec:bpskinputlfmintf} 

\begin{figure}[h]
    \centering
    \includegraphics[width=0.48\textwidth]{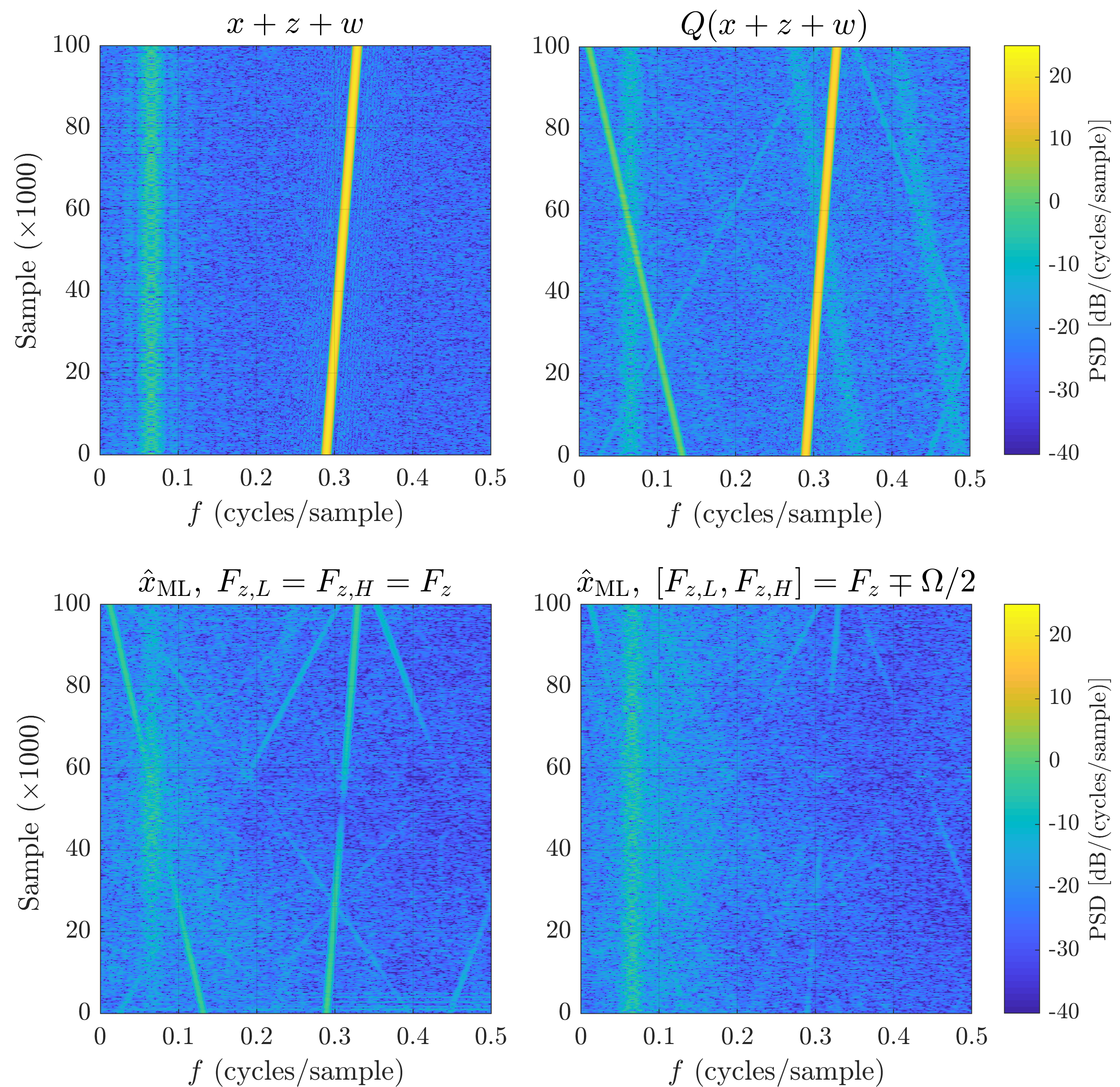}
    \caption{BPSK+LFM Estimator Spectrogram Comparison ($A=0.125, F = 1/16 + \pi/1000, \tau=50, A_{\mathrm{z}} = 1.25, F_{\mathrm{z}} = 5/16 - \pi/1000, \Omega_{\mathrm{z}} = 1/25, \Upsilon_{\mathrm{z}}=10^{5}, N=8$)}
    \label{fig:label17}
\end{figure}

Due to the time-varying (non-stationary) nature of this input we compare its spectrogram in Fig.~\ref{fig:label17} to evaluate the performance of our estimator. The high-power interfering LFM signal saturates the quantizer, creating a third harmonic that aliases into the bandwidth of our desired BPSK signal. The resultant sweep produces bit errors when in-band and reduces SNR by attenuating the BPSK signal when out-of-band. Neither of these effects can be directly compensated with linear filtering. Using a narrowband estimator achieves a limited correction, effective only during a short duration of the LFM sweep when its instantaneous frequency aligns with the prior expectation. A wideband estimator succeeds in achieving consistent cancellation over time as well as restoring BPSK power as desired.

\begin{figure}[h]
    \centering
    \includegraphics[width=0.48\textwidth]{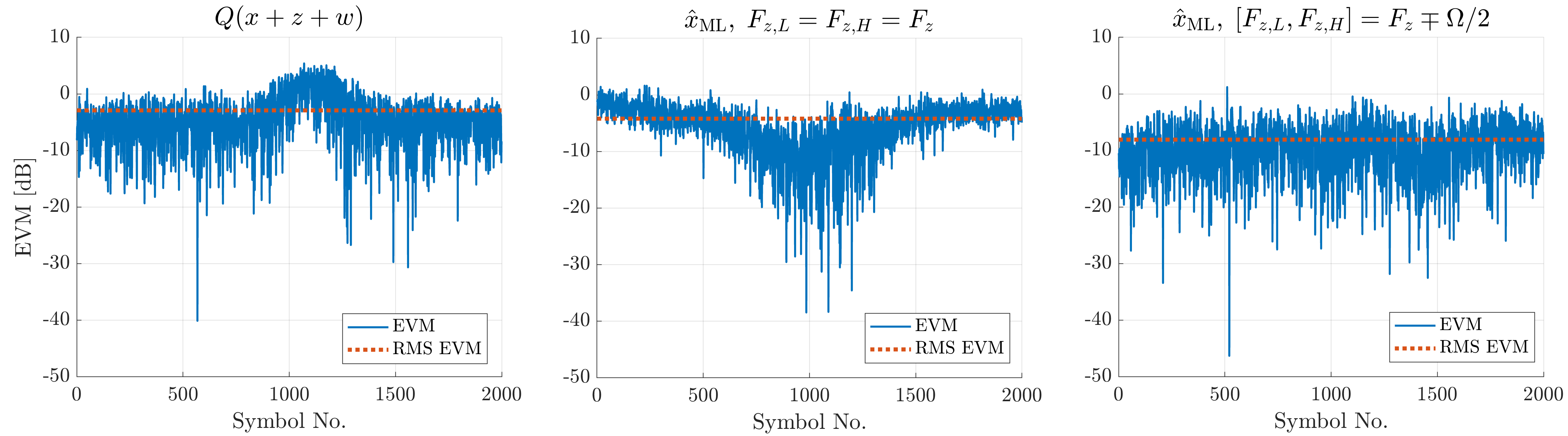}
    \caption{BPSK+LFM Estimator EVM Over Time ($A=0.125, F = 1/16 + \pi/1000, \tau=50, A_{\mathrm{z}} = 1.25, F_{\mathrm{z}} = 5/16 - \pi/1000, \Omega_{\mathrm{z}} = 1/25, \Upsilon_{\mathrm{z}}=10^{5}, N=8$)}
    \label{fig:label18}
\end{figure}

The effects qualitatively described in the previous paragraph are quantitatively justified by the EVM plots over time, given in Fig.~\ref{fig:label18}.

\section{Conclusion}
Three estimators were proposed and evaluated for estimation of the instantaneous desired input in the presence of noise, interference, and low-resolution quantization. In each of five distinct spectral scenarios, the Maximum-Likelihood estimator was shown to achieve comparable or superior results to the Minimum-Mean Square Error and Maximum A-Posteriori Estimators in most metrics of interest with significantly reduced computational overhead. All proposed estimators were shown to be highly effective in reducing both out-of-band interference power and in-band harmonic distortion. They do this while improving MSE and SFDR of the desired signal to levels that would be impossible to recover with traditional linear filtering techniques.


\appendices

\section{General Conditional Distribution Derivation}\label{sec:pyvcondx0derivation}
We can write the general expression for the conditional distribution as:
\begin{align}\label{eq:pyvcondx0_v1}
    p(\mathbf{y} | x_{0}) &= \idotsint_{\mathbb{R}^{U}} \idotsint_{\mathbb{R}^{K}} p(\boldsymbol{\kappa}, \boldsymbol{\mu} | x_{0}) \nonumber\\
    &\hspace{35mm}\cdot p(\mathbf{y} | \boldsymbol{\kappa}, \boldsymbol{\mu}, x_{0}) d\boldsymbol{\kappa}\,d\boldsymbol{\mu} \\
    &= \idotsint_{\mathbb{R}^{U}} p(\boldsymbol{\mu} | x_{0}) \cdot \idotsint_{\mathbb{R}^{K}} p(\boldsymbol{\kappa} | \boldsymbol{\mu}, x_{0}) \nonumber\\
    &\hspace{15mm}\cdot \left(\prod_{n=-N+1}^{0} p(y_{n} | \boldsymbol{\kappa}, \boldsymbol{\mu}, x_{0})\right) d\boldsymbol{\kappa}\,d\boldsymbol{\mu} 
\end{align}
Recall that $\mathbf{x}(\boldsymbol{\kappa})$ and $\mathbf{z}(\boldsymbol{\mu})$ are assumed to be independent meaning $p(\boldsymbol{\mu} | x_{0}) = p(\boldsymbol{\mu})$ and $p(\boldsymbol{\kappa} | \boldsymbol{\mu}, x_{0}) = p(\boldsymbol{\kappa} | x_{0})$. Due to conditional independence $p(y_{n} | \boldsymbol{\kappa}, \boldsymbol{\mu}, x_{0}) = p(y_{n} | \boldsymbol{\kappa}, \boldsymbol{\mu})$. Substituting gives the final expression used in (\ref{eq:pyvcondx0_final}).

To compute (\ref{eq:pyncondku}) we use (\ref{eq:quantizerdefn}) to express:
\begin{multline}
    p(y_{n}| \boldsymbol{\kappa}, \boldsymbol{\mu}) = p(T_{y_{n}} < x_{n}(\boldsymbol{\kappa}) + z_{n}(\boldsymbol{\mu}) + w < T_{y_{n}+1})\\
    = p(T_{y_{n}} - x_{n}(\boldsymbol{\kappa}) - z_{n}(\boldsymbol{\mu}) < w < T_{y_{n}+1} - x_{n}(\boldsymbol{\kappa}) - z_{n}(\boldsymbol{\mu}))\\
    = \int_{T_{y_{n}} - x_{n}(\boldsymbol{\kappa}) - z_{n}(\boldsymbol{\mu})}^{T_{y_{n}+1} - x_{n}(\boldsymbol{\kappa}) - z_{n}(\boldsymbol{\mu})} p(w) dw 
\end{multline}

\section{BPSK Conditional Distribution Derivation}\label{sec:BPSKconddistder}
Prior conditional distributions for $A$ and $F$ are given by (\ref{eq:pacondx0}) and (\ref{eq:pfcondax0}) respectively. 
\begin{equation}
    p(\boldsymbol{\theta} | f, a, x_{0}) = \int_{\mathbb{R}} p(\phi | f, a, x_{0}) \cdot p(\boldsymbol{\theta} | \phi, f, a, x_{0}) d\phi
\end{equation}
The prior conditional distribution for $\Phi$ is given by (\ref{eq:pphicondfax0}) since $\theta_{0} = \phi$ by definition in (\ref{eq:Thetandefn}).

Consider the set $\mathcal{S}$ containing the ensemble of all binary sequences where $\boldsymbol{S_{[i]}}$ is one such sequence and $S_{[i], j}$ is the index-$j$ entry in the $i$-th sequence of the ensemble. Consider $\mathcal{S}' \subset \mathcal{S}$ such that $\mathcal{S}' = \left\{ \boldsymbol{S_{[i]}} \in \mathcal{S} | S_{[i],0} = 0\right\}$ which is the set containing the ensemble of all binary sequences with $0$ value for index-$0$.
\begin{align}
    p(\boldsymbol{\theta} | \phi, f, a, x_{0}) &= \frac{1}{\tau} \sum_{l = 0}^{\tau - 1} p(\boldsymbol{\theta} | l, \phi, f, a, x_{0})\\
    p(\boldsymbol{\theta} | l, \phi, f, a, x_{0}) &\nonumber\\
    &\hspace{-1.8cm}= \frac{1}{|\mathcal{S}'|} \sum_{i \in \mathcal{S}'} \prod_{n} \delta \left(\theta_{n} - \left( \phi + \pi \cdot S'_{[i], \lfloor (n + l)/\tau \rfloor} \right)\right) 
\end{align}

\section{Probability of MPSK Tone Assumption Validity}\label{sec:toneassumptionprobderivation}
A LUT using a sliding window of length $N$ samples is applied to an MPSK signal. Each phase transition point in the MPSK signal is separated by $\tau$ samples. Consequently, at any given instance the LUT window will capture a fixed number of phase transition points which we denote $\bar{N}$. Direct analysis allows us to determine the value of $\bar{N}$ and the proportion of windows that will produce that value since over time the LUT window will slide over each point in the MPSK signal.
\begin{equation}\label{eq:Nbardefn}
    \bar{N} = \left\{\begin{matrix}
    \lceil N/\tau \rceil - 1 & \mathrm{w.p.}\;\; 1 - \frac{N-1}{\tau} + (\lceil N/\tau \rceil - 1)\\
    \lceil N/\tau \rceil & \mathrm{w.p.}\;\; \frac{N-1}{\tau} - (\lceil N/\tau \rceil - 1)
    \end{matrix}\right. 
\end{equation}
where w.p. denotes ``with proportion''. For the tone assumption to be accurate, each symbol within the LUT window must be the same. Each phase transition point in a balanced MPSK signal has a $1/M$ probability of maintaining the same phase as the previous region. Consequently the total probability that all phase transitions align to maintain the tone assumption is:
\begin{multline}
    p\left(\boldsymbol{\Theta} = \Phi\right) = \mathbb{E}\left[(1/M)^{\bar{N}}\right]\\
    = (1/M)^{(\lceil N/\tau \rceil - 1)} \cdot \left(1 - \frac{N-1}{\tau} + (\lceil N/\tau \rceil - 1)\right)\\
    + (1/M)^{(\lceil N/\tau \rceil)} \cdot \left(\frac{N-1}{\tau} - (\lceil N/\tau \rceil - 1)\right)
\end{multline}
Simplifying algebraically gives us (\ref{eq:probToneMPSK}).


\section{BPSK Input Exact Estimator Derivation}\label{sec:BPSKexactconddist}
Consider the lower-triangular square matrix $\mathbf{E}$ defined as:
\begin{equation}
    E_{[i,j]} = \left\{\begin{matrix}
        1, i > j\\
        0, i \leq j\end{matrix}\right.
\end{equation}

This matrix contains the ensemble of possible bit sequences (with oversampling) capturing at most one phase transition in a window of length $N$. Each row vector $\mathbf{e_{[i]}}$ represents one such sequence from time-index $-N+1$ to $0$, with $E_{[i, N]} = 0$ ensuring the $\theta_{0} = \phi$ condition is always met.

\begin{multline}
    p(\boldsymbol{\theta} | \phi, f, a, x_{0}) = \sum_{i = 1}^{N} p(\mathbf{e_{[i]}})\\
    \cdot \prod_{n = -N+1}^{0} \delta \left(\theta_{n} - \left(\phi + \pi \cdot E_{[i,n+N]}\right)\right)
\end{multline}

The first row of $\mathbf{E}$ represents the case for no phase transition (since $\forall j, E_{[1, j]} = 0$) the probability of which is given by (\ref{eq:NlTprobvalidtoneBPSK}). The other $N-1$ sequences in the remaining rows of $\mathbf{E}$ are all equally likely. This gives:

\begin{align}
    p(\mathbf{e_{[i]}}) = \left\{\begin{matrix}
        1 - (N-1)/(2\tau),& i = 1\\
        1/(2\tau),& i = \left\{2, \cdots, N\right\} \end{matrix}\right.
\end{align}

%





\ifCLASSOPTIONcaptionsoff
  \newpage
\fi



\bibliography{refs/References}
\bibliographystyle{IEEEtran}
\end{document}